\documentclass[aps,prc,twocolumn,superscriptaddress,preprintnumbers,showpacs,floatfix,nofootinbib]{revtex4-1}
\usepackage{amsmath,graphicx,color,ulem}
\usepackage{hyperref}
\newcommand{\trento}{T$\mathrel{\protect\raisebox{-2.1pt}{R}}$ENTo}

\begin{document}

\title{Impact parameter dependence of anisotropic flow: Bayesian reconstruction in ultracentral nucleus-nucleus collisions}

\author{Mubarak Alqahtani}
\affiliation{Department of Physics, College of Science, Imam Abdulrahman Bin Faisal University, Dammam 31441, Saudi Arabia  }
\affiliation{Basic and Applied Scientific Research Center, Imam Abdulrahman Bin Faisal University, Dammam 31441 Saudi Arabia}

\author{Rajeev S. Bhalerao}
\affiliation{Department of Physics, Indian Institute of Science Education and Research (IISER), Homi Bhabha Road, Pune 411008, India}

\author{Giuliano Giacalone}
\affiliation{Theoretical Physics Department, CERN, 1211 Geneva 23, Switzerland}
\affiliation{Institut f\"ur Theoretische Physik, Universit\"at Heidelberg, Philosophenweg 16, 69120 Heidelberg, Germany} 

\author{Andreas Kirchner}
\affiliation{Institut f\"ur Theoretische Physik, Universit\"at Heidelberg, Philosophenweg 16, 69120 Heidelberg, Germany} 

\author{Jean-Yves Ollitrault}
\affiliation{Institut de physique th\'eorique, Universit\'e Paris Saclay, CNRS, CEA, F-91191 Gif-sur-Yvette, France}

\date{\today}

\begin{abstract}
Peculiar phenomena have been observed in analyses of anisotropic flow ($v_n$) fluctuations in ultracentral nucleus-nucleus collisions:
The fourth-order cumulant of the elliptic flow ($v_2$) distribution changes sign.    
In addition, the ATLAS collaboration has shown that cumulants of $v_n$ fluctuations of all orders depend significantly on the centrality estimator. 
We show that these peculiarities are due to the fact that the impact parameter $b$ always spans a finite range for a fixed value of the centrality estimator. 
We provide a quantitative determination of this range through a simple Bayesian analysis.  
We obtain excellent fits of STAR and ATLAS data, with a few parameters, by assuming that the probability distribution of $v_n$ solely depends on $b$ at a given centrality. 
This probability distribution is almost Gaussian, and its parameters depend smoothly on $b$, in a way that is constrained by symmetry and scaling laws. 
We reconstruct, thus, the impact parameter dependence of the mean elliptic flow in the reaction plane in a model-independent manner, and assess the robustness of the extraction using Monte Carlo simulations of the collisions where the impact parameter is known.
We argue that the non-Gaussianity of $v_n$ fluctuations gives direct information on the hydrodynamic response to initial anisotropies, ATLAS data being consistent with a smaller response for $n=4$ than for $n=2$ and $n=3$, in agreement with hydrodynamic calculations.  
\end{abstract}

\preprint{CERN-TH-2024-118}

\maketitle
\section{Introduction}
There is growing interest in the study of ultracentral nucleus-nucleus collisions at ultrarelativistic energies. 
Their impact parameter is close to zero, which results in average geometries that are nearly isotropic in the transverse plane~\cite{Luzum:2012wu}. 
Ultra-central collisions have been studied in dedicated analyses~\cite{CMS:2013bza} (and by means of specific triggers~\cite{Aaboud:2018ves}) of Pb+Pb collisions at the Large Hadron Collider (LHC). 
One of the well-known outputs of these analyses is a tension between theory and experiment, referred to as the  ``ultracentral flow puzzle'':
The measured triangular flow, $v_3$, is larger than expected relative to elliptic flow, $v_2$~\cite{Luzum:2012wu}. 
This discrepancy has been addressed in a number of theoretical studies~\cite{Shen:2015qta,Alba:2017hhe,Carzon:2020xwp,Nijs:2021clz,Giannini:2022bkn,Kuroki:2023ebq}, and the puzzle is not yet fully resolved. 
Recently, ultracentral collisions have also been used as a laboratory to measure the compressibility, or speed of sound, of the quark-gluon plasma~\cite{Gardim:2019brr,Nijs:2023bzv,CMS:2024sgx,Gardim:2024zvi,SoaresRocha:2024drz,Sun:2024zsy,ATLAS:2024jvf}, and to highlight the impact of the ground-state deformation of the colliding ions on the collective flow \cite{STAR:2015mki,ALICE:2018lao,ALICE:2021gxt,ATLAS:2022dov,STAR:2024eky}.

Another interesting observation in ultracentral collisions concerns the fourth-order cumulant of the elliptic flow distribution, denoted by $c_2\{4\}$, and measured as a four-particle correlation, which changes sign from negative to positive as the collisions become more central. 
The negative value is well understood~\cite{Borghini:2000sa} and is generated by elliptic flow in the reaction plane. 
In central collisions, the distribution of the anisotropic flow vector is dominated by fluctuations~\cite{PHOBOS:2006dbo} which are essentially Gaussian~\cite{Voloshin:2007pc}, implying that all cumulants should vanish beyond second order (variance). 
Therefore, one would expect $c_2\{4\}$ to go smoothly from negative to zero as the centrality percentile decreases. 
While it is understood that this behavior is associated with the capability of the experiments to resolve the  \textit{centrality} of the collisions \cite{Zhou:2018fxx}, we still miss a quantitative understanding of how an imperfect determination of the collision impact parameter (the \textit{true centrality} variable) generates
the {\it positive\/} $c_2\{4\}$ values measured in Au+Au collisions at RHIC~\cite{STAR:2015mki}, and in Pb+Pb collisions at the LHC~\cite{ATLAS:2019peb}. In turn, one could address whether the same physical mechanism explains the observation of the ATLAS collaboration that the root-mean-square value of $v_n$, $v_n\{2\}$, differs by a few per cent, still in ultracentral collisions,  depending on whether one defines the collision centrality using the charged multiplicity ($N_{\rm ch}$) or the transverse energy ($E_T$).

In this paper, we show that, indeed, both observations related to $v_n\{2\}$ and $c_2\{4\}$ are naturally explained if one properly takes into account the difference between the centrality as defined experimentally (using typically the multiplicity in some detector~\cite{ALICE:2013hur}) and the true centrality~\cite{Das:2017ned}, as defined by the impact parameter, $b$, of the collision. 
The impact parameter determines the geometry of the collision area. 
It plays a key role not only for elliptic flow, but more generally for any phenomenon involving collective flow~\cite{Samanta:2023amp}.
Its probability distribution at fixed multiplicity can be precisely reconstructed using a simple Bayesian analysis~\cite{Das:2017ned}, which is recalled in Sec.~\ref{s:centrality}.  

Assuming that the probability distribution of the anisotropic flow vector is close to a two-dimensional Gaussian \cite{Voloshin:2007pc} that depends solely on $b$, in Sec.~\ref{s:data} we fit experimental data on the cumulants of flow fluctuations (whose definitions are recalled in Sec.~\ref{s:notations}) measured by the ATLAS and the STAR collaborations (as presented in Sec.~\ref{s:expresults}). 
From purely Gaussian fits (Sec.~\ref{s:gaussian}), we achieve in Sec.~\ref{s:centralitydependence}  a reconstruction of the impact parameter dependence of the mean elliptic flow in the reaction plane, and of the width of $v_n$ fluctuations, for $n=2,3,4$. 
The change of sign of the fourth-order cumulant of the elliptic flow distribution is then discussed in Sec.~\ref{s:c24}.
We show that implementing leading non-Gaussian corrections (discussed in Sec.~\ref{s:nongaussian}), which are the skewness (for elliptic flow) and kurtosis (for all other harmonics), in the fit to data improves the description of the experimental results.  
We argue that the non-Gaussianity of the $n$th flow harmonic coefficient distribution gives direct information about the hydrodynamic response coefficient, $v_n/\varepsilon_n$, where $\varepsilon_n$ denotes the spatial anisotropy of the initial entropy density profile \cite{Teaney:2010vd}. 

In Sec.~\ref{s:trento}, we test our assumptions against models where the impact parameter is known. 
Since $v_n$ stems from linear response to $\varepsilon_n$~\cite{Ollitrault:1992bk,Alver:2010gr,Teaney:2010vd}, we need a Monte Carlo model of initial entropy density fluctuations. 
We use the popular \trento{} model~\cite{Moreland:2014oya}. 
We point out that the distribution of $\varepsilon_n$ at fixed $b$ is somewhat correlated with the event multiplicity. 
This, in turn, implies that our starting assumption, that the distribution of $v_n$ depends solely on $b$, is not quite correct. 
We estimate the impact of this correlation on our results.

In Sec.~\ref{s:conclusions}, we summarize our results and discuss avenues for research opened by our analysis, arguing in particular for future hydrodynamic calculations at fixed impact parameter to facilitate theory-to-data comparisons.

Technical material related to cumulant expansions is relegated to Appendix \ref{s:cumulants}.

\section{Centrality fluctuations}
\label{s:centrality}

\begin{figure*}[ht]
\begin{center}
\includegraphics[width=.9\linewidth]{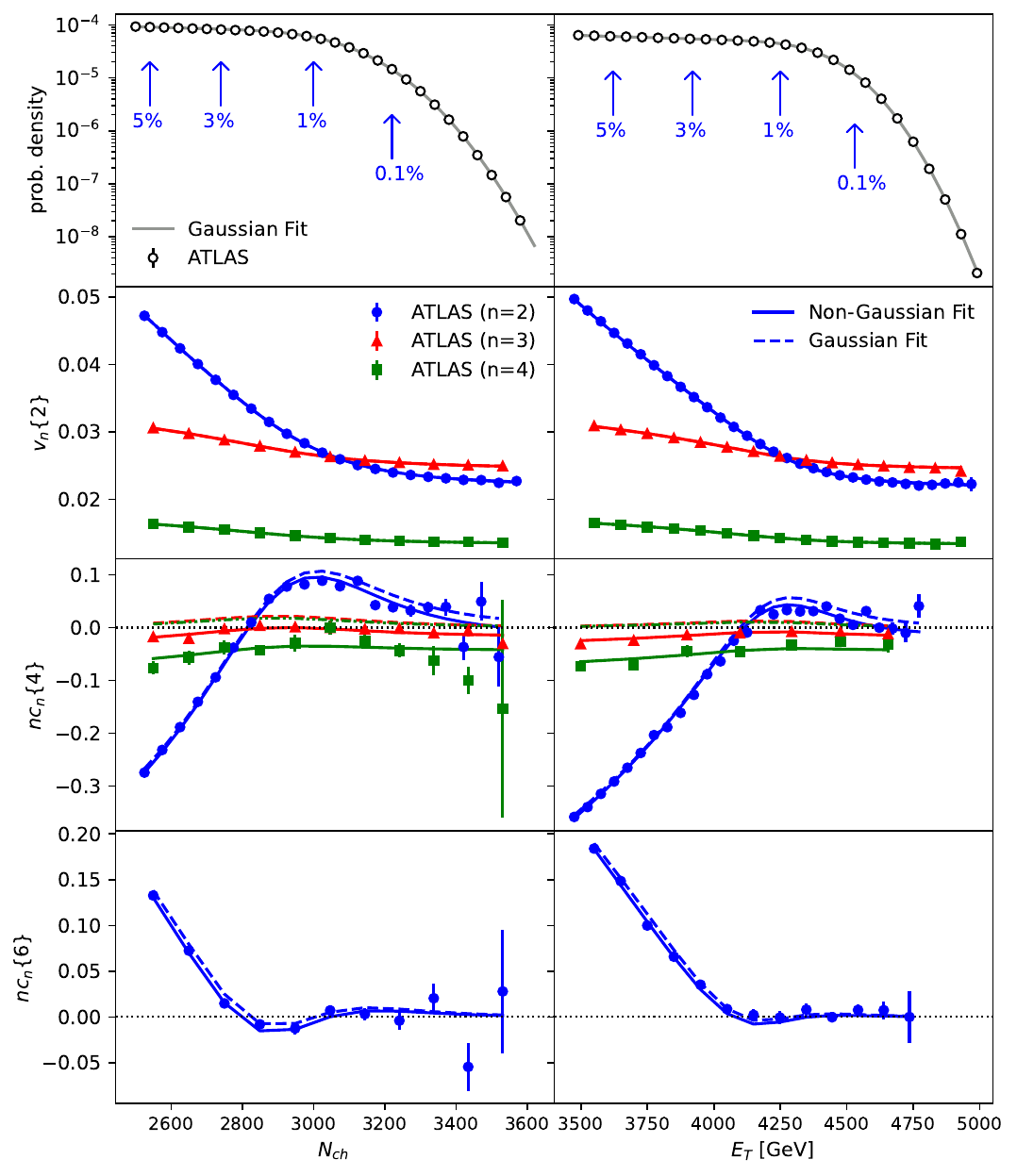} 
\end{center}
\caption{ 
  \label{fig:ATLAS}
  Top: distributions of the charged multiplicity $N_{ch}$ (left) and transverse energy $E_T$ (right) measured by the ATLAS collaboration in Pb+Pb collisions at $\sqrt{s_{NN}}=5.02$~TeV~\cite{ATLAS:2019peb}.
  Lines are fits using Eqs.~(\ref{gaussian}) and (\ref{fitpn})~\cite{Yousefnia:2021cup}, and 
  arrows label selected values of the experimentally-defined centrality $c_{\rm exp}$ (see text). 
Lower panels display the cumulants of anisotropic flow, defined by Eqs.~(\ref{defcumulants}) and (\ref{defncumulants}), measured in events with given $N_{ch}$ (left) or $E_T$ (right), for charged particles in the transverse momentum range $0.5<p_t<5$~GeV/$c$.
Dashed lines are fits assuming Gaussian flow fluctuations at fixed impact parameter (Sec.~\ref{s:gaussian}) and solid lines are fits with non-Gaussian corrections added (Sec.~\ref{s:nongaussian}).
Dashed lines and solid lines are on top of one another for $v_n\{2\}$, so that only the solid lines appear. 
  }
\end{figure*}    

\begin{figure}[ht]
\begin{center}
\includegraphics[width=\linewidth]{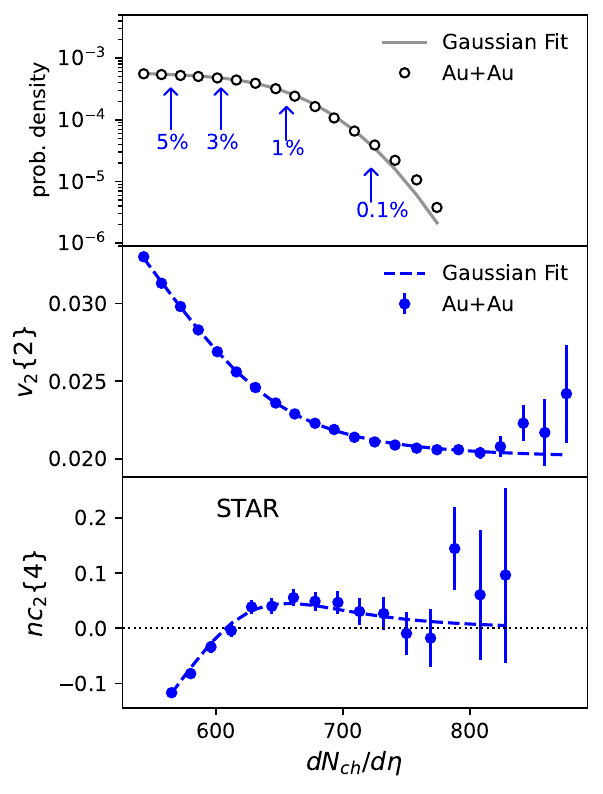} 
\end{center}
\caption{ 
  \label{fig:STAR}
  STAR results on Au+Au collisions at $\sqrt{s_{NN}}=200$~GeV~\cite{STAR:2015mki}. Top: Probability distribution of $dN_{\rm ch}/d\eta$ in Au+Au. The line is a fit using Eqs.~(\ref{gaussian}) and (\ref{fitpn}).
As in Fig.~\ref{fig:ATLAS}, arrows label some values of the experimentally-defined centrality $c_{\rm exp}$. 
  Middle and bottom: $v_2\{2\}$ and $nc_2\{4\}$, defined from Eqs.~(\ref{defcumulants}) and (\ref{defncumulants}), for charged particles in the transverse momentum range $0.2<p_t<2$~GeV/$c$. 
  Lines are fits assuming Gaussian flow fluctuations at fixed impact parameter (Sec.~\ref{s:gaussian}).
  }
\end{figure}    

We define the centrality $c$ of a collision from the cumulative distribution function of the minimum bias impact parameter distribution. 
Typically for $c\lesssim 75\%$, it is related to $b$ by the geometric relation: 
\begin{equation}
\label{defcb}
c=\frac{\pi b^2}{\sigma_{\rm inel}},
\end{equation}
where $\sigma_{\rm inel}$ is the inelastic nucleus-nucleus cross section (which is $767\pm26$ fm$^2$ in Pb+Pb collisions at LHC~\cite{ALICE:2022xir}, and about 685 fm$^2$ for Au+Au collisions at RHIC). 
We only study collisions central enough that Eq.~(\ref{defcb}) holds. 
All our results will be given in terms of $c$, but Eq.~(\ref{defcb}) can be used to express them in terms of $b$. 

The centrality of a collision, as defined by Eq.~(\ref{defcb}), is not measured.  
In heavy-ion experiments, one uses a specific observable as a proxy for the centrality, typically the charged hadron multiplicity or transverse energy collected in a detector. 
We denote this observable generically by $N$.  
The distribution of $N$, which we denote by $P(N)$, is displayed in the top rows of Figs.~\ref{fig:ATLAS} and \ref{fig:STAR} for ATLAS data (Pb+Pb collisions at $\sqrt{s_{NN}}=5.02$~TeV) and STAR data (Au+Au collisions at $\sqrt{s_{NN}}=200$~GeV), respectively.  
Arrows label selected values of the experimentally-defined centrality $c_{\rm exp}$, which is the tail distribution (complementary cumulative distribution function) of $N_{ch}$ or $E_T$, depending on which of these two centrality estimators one chooses.

In this section, we briefly recall how the probability distribution of $c$ at fixed $N$, $P(c|N)$, can be reconstructed from $P(N)$~\cite{Das:2017ned}.\footnote{Note that in Ref.~\cite{Das:2017ned}, the centrality was denoted by $c_b$.}

The idea is to first determine the probability of $N$ at fixed centrality, $P(N|c)$ and then use Bayes' theorem, as we recall below. 
Model calculations show that, for the collision systems under consideration, the distribution of $N$ at fixed $c$ is Gaussian to a very good approximation: 
\begin{equation}
\label{gaussian}
  P(N|c)=\frac{1}{\sqrt{2\pi}\sigma_N(c)}\exp\left(-\frac{(N-\bar N(c))^2}{2\sigma_N^2(c)}\right),
\end{equation}
where $\bar N(c)$ and $\sigma_N^2(c)$ denote the mean and variance of $N$, which depend on centrality.
Throughout this paper, we use the notation $\langle f|c\rangle$ to denote the average value of $f$ at fixed $c$.
With this notation, $\bar N(c)$ and $\sigma_N^2(c)$ can be written in the following way: 
\begin{eqnarray}
  \label{defnbarsigman}
  \bar N(c)&\equiv& \langle N|c\rangle,\cr
  \sigma_N^2(c)&\equiv&  \langle N^2|c\rangle-\langle N|c\rangle^2. 
\end{eqnarray}

The measured distribution $P(N)$ is the integral over centrality: 
\begin{equation}
  \label{fitpn}
P(N)=\int_0^1P(N|c)dc.
\end{equation}
By fitting Eqs.~(\ref{gaussian}) and (\ref{fitpn}) to the measured $P(N)$, one can reconstruct $\bar N(c)$ (assumed to be a positive analytic function, typically the exponential of a polynomial)  and $\sigma_N^2(c=0)$~\cite{Das:2017ned}.

Note that the centrality dependence of $\sigma_N^2(c)$ cannot be inferred from data~\cite{Das:2017ned}, so that assumptions must be made.
Since we are only interested in ultracentral collisions with $c\ll 1$, $\sigma_N^2(c)$ is likely to differ little from $\sigma_N^2(0)$ anyway, and we expect results to be robust, irrespective of the centrality dependence of $\sigma_N^2(c)$. 
In order to quantify the corresponding uncertainty, we test different scenarios. 
In Refs.~\cite{Das:2017ned,Yousefnia:2021cup}, it was assumed that the variance was proportional to the mean. This is the case if the multiplicity is the sum of contributions from identical, independent sources. 
We use a more general parametrization introduced in Ref.~\cite{Samanta:2023kfk},
where the ratio of variance to mean varies linearly with the mean, which is empirically motivated by the results of initial-state models. 
Specifically, we assume that 
\begin{equation}
\label{defgamma}
\frac{\sigma_N^2(c)}{\bar N(c)}=\frac{\sigma_N^2(0)}{\bar N(0)}\left(\gamma + (1-\gamma)\frac{\bar N(c)}{\bar N(0)}\right), 
\end{equation}
where $\gamma$ is a constant. 
The assumption that the variance is proportional to the mean corresponds to $\gamma=1$.  
State-of-the-art models favor a larger value, $\gamma\simeq 2$~\cite{Samanta:2023kfk}. 
This implies that fluctuations are smaller by a factor of two in central collisions than in peripheral collisions, relative to the expectation from independent sources. 
Note that there is at present no direct evidence from data for such a suppression of fluctuations in central collisions. 
The statement is that global theory to data comparisons favor models where such a suppression occurs. 
We vary $\gamma$ between 1 and 3 to study the sensitivity of our results to this parameter. 

Our fits to $P(N)$ are displayed as lines in the top panels of Fig.~\ref{fig:ATLAS} and Fig.~\ref{fig:STAR}. 
For ATLAS data, the difference between fit and data is very small, at the percent level. 
The fit parameters are given in Table I of Ref.~\cite{Yousefnia:2021cup}.
For STAR data, the collaboration does not provide the histogram of $N_{ch}$~\cite{STAR:2015mki}, but only an analytic parametrization of the relation between $N_{ch}$ and the experimentally-defined centrality (see caption of Fig.~\ref{fig:ATLAS}). 
We use this relation to infer the probability distribution of $N_{ch}$, following the procedure described in Ref.~\cite{Giacalone:2018apa}. 
One sees in the upper panel of Fig.~\ref{fig:STAR} that our fit deviates from data in the tail of the distribution.\footnote{For STAR data, we obtain $\bar N(c)=662\exp(-3.2c-0.8c^2-1.4 c^3)$ and $\sigma_N(c=0)= 42$.}
We believe that the reason is that the parametrization used by STAR is only approximate in the tail. 
For this reason, our analysis of STAR data is much less precise than that of ATLAS data. 
Precision could be greatly improved if the histogram of $N_{ch}$ was provided by the STAR collaboration.

\begin{figure}[t]
\begin{center}
\includegraphics[width=\linewidth]{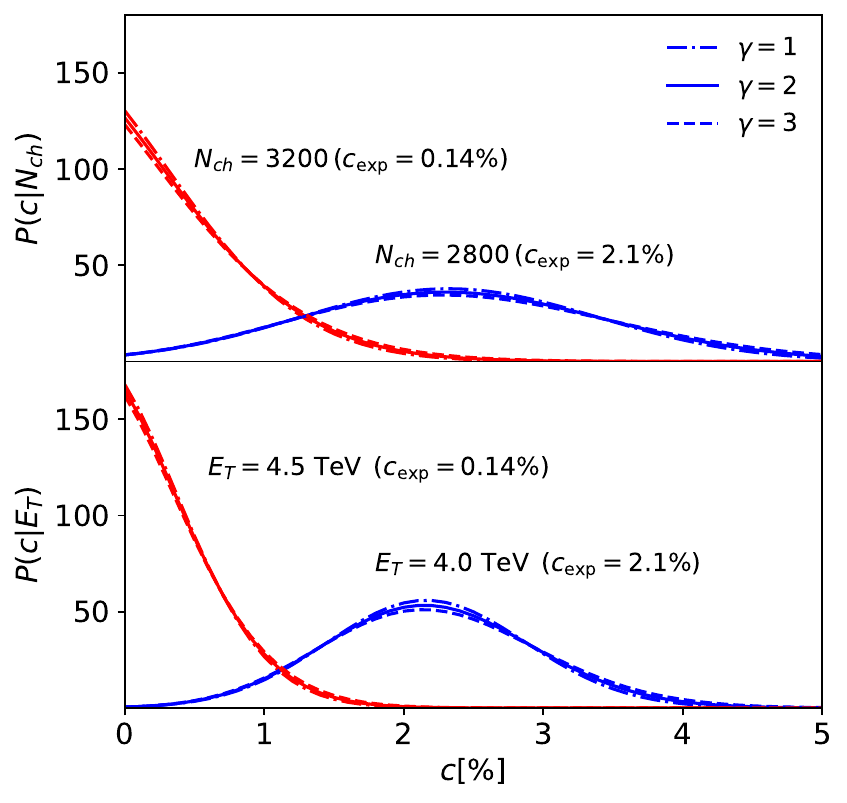} 
\end{center}
\caption{
   \label{fig:probc}
   Distribution of true centrality, $c$, in Pb+Pb collisions at $5.02$~TeV for two values of $N_{ch}$ (top), two values of $E_T$ (bottom), corresponding to two values of the experimentally-defined centrality $c_{\rm exp}$, 
   and three values of the parameter $\gamma$ in Eq.~(\ref{defgamma}). 
 }
\end{figure}    
The fit determines the probability $P(N|c)$, defined by Eq.~(\ref{gaussian}). 
The probability distribution of $c$ at fixed $N$ is then given by Bayes' theorem:\footnote{Note that the distribution of $c$ is flat since it is defined as a cumulative distribution, namely, $P(c)=1$.} 
\begin{equation}
\label{bayes}
P(c|N)=\frac{P(N|c)}{P(N)}.
\end{equation}
An approximate analytic form of $P(c|N)$ can be derived using Eq.~(\ref{gaussian}), expanding $\bar N(c)$ to first order in $c$ around $c=0$, and neglecting the variation of the width, $\sigma_N(c)\approx \sigma_N(0)$. 
A simple calculation then gives:
\begin{equation}
  \label{gaussianc}
P(c|N)\propto \exp\left(-\frac{\left(c-\bar c(N)\right)^2}{2\sigma_c^2}\right),
\end{equation}
where we have omitted the normalization constant, and 
\begin{align}
  \label{evalsigmac}
\bar c(N)&\equiv\frac{\bar N(0)-N}{-(d\bar N/dc)_{c=0}}\\
\sigma_c&\equiv\frac{\sigma_N(0)}{-(d\bar N/dc)_{c=0}}. 
\end{align}
Eq.~(\ref{gaussianc}) shows that the distribution of centrality at fixed $N$ is approximately Gaussian. 

Distributions corresponding to Pb+Pb collisions at 5.02~TeV and to the ATLAS detector calculated numerically using Eq.~(\ref{bayes}) are displayed in Fig.~\ref{fig:probc} for two values of $N_{ch}$ and two values of $E_T$. 
For the smaller values  (blue lines), one sees that the distribution is approximately Gaussian, as expected from Eq.~(\ref{gaussianc}). 
The distribution of $c$ is broader at fixed $N_{ch}$ (top panel) than at fixed $E_T$ (bottom panel), corresponding to a larger  $\sigma_ c$. 
Using the numerical values in Table I of Ref.~\cite{Yousefnia:2021cup}, we indeed obtain $\sigma_c\simeq 1.2\%$ if the centrality estimator is $N_{ch}$, and $\sigma_c\simeq 0.85\%$ if the centrality estimator is $E_T$. 
This means that $E_T$ is a better centrality estimator\footnote{This is due to the different pseudorapidity coverage~\cite{Yousefnia:2021cup}.}  than $N_{ch}$.
For the larger values of $N_{ch}$ or $E_T$ (red lines), the Gaussian is truncated due to the boundary condition $c\ge 0$. 
The cases displayed actually correspond to ultracentral collisions  where $c(N)<0$~\cite{Samanta:2023amp}, implying that the most probable value of $c$ is $0$.

Let us discuss the dependence of results on the value of the parameter $\gamma$. 
Larger $\gamma$ implies larger $\sigma_N^2(c)$, that is, larger fluctuations of $N$ for fixed $c$, which in turn implies larger fluctuations of $c$ for fixed $N$. 
This is confirmed by the numerical results in Fig.~\ref{fig:probc}, where $P(c|N)$ become broader as $\gamma$ increases. However, this effect is nearly negligible. We shall use the value $\gamma=2$ throughout Sec.~\ref{s:data}, where we fit experimental results.

Finally, note that in the case of STAR data, the width of the centrality distribution would be twice as large as for ATLAS data, as also shown in Ref.~\cite{Das:2017ned}.

\section{Understanding ATLAS and STAR data}
\label{s:data}

In this section, we interpret the experimental results on anisotropic flow cumulants displayed in Figs.~\ref{fig:ATLAS} and \ref{fig:STAR}.

\subsection{Definitions and notations}
\label{s:notations}

We first briefly recall the definition of anisotropic flow and of its cumulants. 
The ``flow picture''~\cite{Luzum:2011mm}, which is assumed throughout this paper, is that in a collision event particles are emitted independently according to an underlying probability distribution, $P(\varphi)$, where $\varphi$ is the azimuthal angle.
Throughout this paper, we choose the direction of the impact parameter as the $x$ axis, corresponding to $\varphi=0$. 
The complex~\cite{Gardim:2011xv} anisotropic flow coefficient~\cite{Voloshin:1994mz} is defined as
\begin{eqnarray}
  V_n=
  \int_0^{2\pi}e^{in\varphi}P(\varphi)\, d\varphi. 
\end{eqnarray}
The probability distribution $P(\varphi)$ fluctuates event by event, and so does $V_n$.
The experimentally-measured cumulants of order 2, 4 and 6 are defined by~\cite{ATLAS:2019peb}:
\begin{eqnarray}
  \label{defcumulants}
c_n\{2\} &=& \langle |V_n|^2\rangle=v_n\{2\}^2\cr 
c_n\{4\} &=& \langle |V_n|^4\rangle-2\langle |V_n|^2\rangle^2=-v_n\{4\}^4 \cr 
c_n\{6\} &=& \langle |V_n|^6\rangle-9\langle |V_n|^4\rangle\langle |V_n|^2\rangle+12\langle |V_n|^2\rangle^3\cr
&=& 4v_n\{6\}^6,
\end{eqnarray}
where angular brackets denote an average value over events for a given value of the centrality estimator.  
The ATLAS collaboration normalize cumulants of order four and higher in such a way that they are invariant through a global rescaling of $V_n$, by introducing: 
\begin{eqnarray}
  \label{defncumulants}
  nc_n\{4\} &\equiv &  \frac{c_n\{4\}}{c_n\{2\}^2}=- \frac{v_n\{4\}^4}{v_n\{2\}^4}\cr
  nc_n\{6\} &\equiv &  \frac{c_n\{6\}}{4c_n\{2\}^3}= \frac{v_n\{6\}^6}{v_n\{2\}^6}. 
\end{eqnarray}

\subsection{Presentation of results from ATLAS and STAR}
\label{s:expresults}

The ATLAS collaboration measure $v_n\{2\}$ (rms anisotropic flow) and $nc_n\{4\}$ for $n=2,3,4$, as well as $nc_2\{6\}$, while the STAR collaboration measure $v_2\{2\}$ and $v_2\{4\}^4=-c_2\{4\}$. 
The ATLAS analysis is carried out with two different centrality estimators, the charged multiplicity near mid-rapidity ($N_{\rm ch}$, left panels of Fig.~\ref{fig:ATLAS}) and the transverse energy in calorimeters at larger rapidities ($E_T$, right panels). 

Both ATLAS and STAR analyses are carried out for all centralities (typically from 0 to 80\%).
Our study focuses on central collisions. Therefore, we only show the results for 5\% of events corresponding to the largest values of $N_{ch}$ or $E_T$.

The second-order cumulants yield the rms anisotropic flow: 
$v_3\{2\}$ and $v_4\{2\}$ originate from initial-state fluctuations and depend mildly on $N_{ch}$ or $E_T$, whereas $v_2\{2\}$ shows a strong increase as the multiplicity decreases. The latter observation corresponds to the onset of in-plane elliptic flow, driven by the almond geometry of the overlap area~\cite{Ollitrault:1992bk}.

Higher-order cumulants have been much studied for elliptic flow~\cite{ALICE:2010suc,CMS:2017glf,ALICE:2018rtz}. 
The standard picture~\cite{Voloshin:2007pc} is that $v_2\{4\}$ and $v_2\{6\}$ coincide to a good approximation with the mean elliptic flow in the reaction plane, resulting in a negative $c_2\{4\}$ and positive $c_2\{6\}$ [Eq.~(\ref{defcumulants})]. 
Equation~(\ref{defncumulants}) implies the approximate relation
\begin{equation}
nc_2\{6\}\simeq (-nc_2\{4\})^{3/2}, 
\end{equation}
which holds in off-central collisions. 
For central collisions, or large values of the multiplicity, $nc_2\{4\}$ becomes positive for ATLAS data and, correspondingly, $v_2\{4\}^4$ becomes negative. The same behavior is observed for STAR data.  
A hint of this change of sign had previously been seen by ALICE (see Fig. 4 of Ref.~\cite{ALICE:2014dwt}).
This phenomenon is discussed in detail in Sec.~\ref{s:c24}, where we show that the positive $c_2\{4\}$ is a generic phenomenon, generated by the centrality fluctuations discussed in Sec.~\ref{s:centrality}. 

The higher-order cumulants of $v_3$~\cite{ALICE:2011ab} and $v_4$~\cite{ATLAS:2014qxy} have been less studied, though they can be measured very accurately. The negative values measured by the ATLAS collaboration for $nc_3\{4\}$ and $nc_4\{4\}$ in Fig.~\ref{fig:ATLAS} imply in particular non-Gaussian flow fluctuations, which will be discussed in Sec.~\ref{s:nongaussian}.

\subsection{Gaussian fluctuations}
\label{s:gaussian}

We now introduce our model, which will be further refined in Sec.~\ref{s:nongaussian}. 
We assume that the probability distribution of $V_n$ at fixed centrality is approximately Gaussian~\cite{Voloshin:2007pc}, which amounts to applying the central-limit theorem to flow fluctuations.  
Dropping the subscript $n$ for simplicity, and separating the complex $V$ into real and imaginary parts, $V=v_x+iv_y$, we write: 
\begin{equation}
  \label{gaussianv}
  P(V|c)=\frac{1}{\pi \sigma_v^2(c)}\exp\left(-
  \frac{\left(v_x-\bar v(c)\right)^2+v_y^2}{\sigma_v^2(c)}\right), 
\end{equation}
where $\bar v(c)$ and $\sigma_v^2(c)$ denote the mean and variance of $V$ at fixed $c$:
\begin{eqnarray}
  \label{defvbarsigmav}
  \bar v(c)&\equiv& \langle V|c\rangle\cr
  \sigma_v^2(c)&\equiv&  \langle |V|^2|c\rangle-|\langle V|c\rangle|^2. 
\end{eqnarray}
Note the similarity of Eqs.~(\ref{gaussianv}) and (\ref{defvbarsigmav}) with Eqs.~(\ref{gaussian}) and (\ref{defnbarsigman}).
We recall that a strong nuclear deformation spoils the Gaussian approximation~\cite{Giacalone:2018apa}, and that we assume spherical nuclei throughout this paper.\footnote{This is also the reason why we do not analyze STAR data on U+U collisions \cite{STAR:2015mki} in this work.}

The mean, $\bar v(c)$, is along the $x$ axis because the average collision geometry (obtained from an ensemble of events with the same impact parameter) is symmetric under $y\to -y$ (symmetry with respect to the reaction plane).
For symmetric collisions at midrapidity and $c>0$, symmetry under $\varphi\to\varphi+\pi$ implies that $\bar v(c)$ vanishes for $v_3$. 
For collisions at $c=0$ (zero impact parameter), azimuthal symmetry implies that $\bar v(c)$ vanishes also for even harmonics.

In practice, we consider a non-zero $\bar v(c)$ only for elliptic flow, where it corresponds to the mean elliptic flow in the reaction plane~\cite{Ollitrault:2023wjk}. 
$\bar v(c)$ is an analytic function of $c$ which vanishes at $c=0$. We expand it in powers of the centrality:  
\begin{equation}
\label{polynom1}
\bar v(c)=a_1c+a_2c^2+a_3c^3. 
\end{equation}
Higher orders in $c$ are not needed for the range of centralities considered in this paper. 
For quadrangular flow, $v_4$, the leading contribution to $\bar v(c)$ allowed by symmetry is proportional to $c^2$, corresponding to $v_4\sim v_2^2$~\cite{Borghini:2005kd}, and we have checked that it is negligible in the centrality range that we consider in this paper.

The variance, $\sigma_v^2(c)$, is also an analytic function of $c$.
For each harmonic, we expand it in powers of $c$:
\begin{equation}
\label{polynom2}
\sigma_v^2(c)=A_0+A_1c+A_2c^2.
\end{equation}
Again, three terms suffice for the considered centrality range.

The experimentally-measured cumulants  in Eq.~(\ref{defcumulants}) are functions of moments $\langle |V|^k\rangle$ (we have again dropped the subscript $n$ for simplicity), where the average value is taken at fixed $N$. Thus, we introduce the more appropriate notation $\langle |V|^k|N\rangle\equiv\langle |V|^k\rangle$.
The crucial point is that each $N$ corresponds to a range of centralities, as discussed in Sec.~\ref{s:centrality}.
Therefore, these moments are evaluated in two steps. 
First, one evaluates them at fixed centrality. Then, one averages over centralities.

In our approach, the moments at fixed centrality are those of the Gaussian distribution  in Eq.~(\ref{gaussianv}):
\begin{eqnarray}
  \label{gaussmoments}
  \langle |V|^2|c\rangle&=& \sigma_v^2+\bar v^2\cr
  \langle |V|^4|c\rangle&=&2\sigma_v^4+4\sigma_v^2\bar v^2+
  \bar v^4  \cr
  \langle |V|^6|c\rangle&=&6\sigma_v^6+18\sigma_v^4\bar v^2+9\sigma_v^2\bar v^4+
  \bar v^6,   
\end{eqnarray}
where we have used the shorthand notations $\bar v$ and $\sigma_v^2$ for  $\bar v(c)$ and $\sigma_v^2(c)$.

The average over centrality is done using the probability $P(c|N)$ defined in Sec.~\ref{s:centrality}:
\begin{equation}
\label{particular}
  \langle |V|^k|N\rangle=\int \langle |V|^k|c\rangle  P(c|N)dc. 
\end{equation}
Finally, from these moments one evaluates the cumulants in Eq.~(\ref{defcumulants}).

For each harmonic, we carry out a combined fit of experimental data with this model.
In the case of ATLAS data, the ``combined'' fit uses both $N_{ch}$- and $E_T$-binned data, and all cumulant orders.
For $n=3$ and $n=4$, there are 3 fit parameters, which are the $A_i$ in Eq.~(\ref{polynom2}).
For $n=2$, there are 3 additional fit parameters, namely, the $a_i$ in Eq.~(\ref{polynom1}).   

The fits are displayed as dashed lines in Figs.~\ref{fig:ATLAS} and \ref{fig:STAR}.
They are excellent for $v_n\{2\}$. In Fig.~\ref{fig:ATLAS}, this implies that differences in rms $v_n$ results due to the different centrality determination method, as emphasized by the ATLAS collaboration \cite{ATLAS:2019peb}, are fully captured in our approach. In the case of elliptic flow, $v_2$, our fits explain as well the change of sign of fourth-order cumulant,  $c_2\{4\}$, again capturing the differences due to the centrality determination between the left and right panels of Fig.~\ref{fig:ATLAS}.
For triangular and quadrangular flows, $v_3$ and $v_4$, the model can not reproduce the measured fourth-order cumulants. 
For this reason, non-Gaussian corrections must be included. 
They will be discussed in Sec.~\ref{s:nongaussian}, where the values of the fit parameters are also listed.   

\subsection{The centrality dependence of anisotropic flow}
\label{s:centralitydependence}

\begin{figure*}[t]
\begin{center}
\includegraphics[width=.9\linewidth]{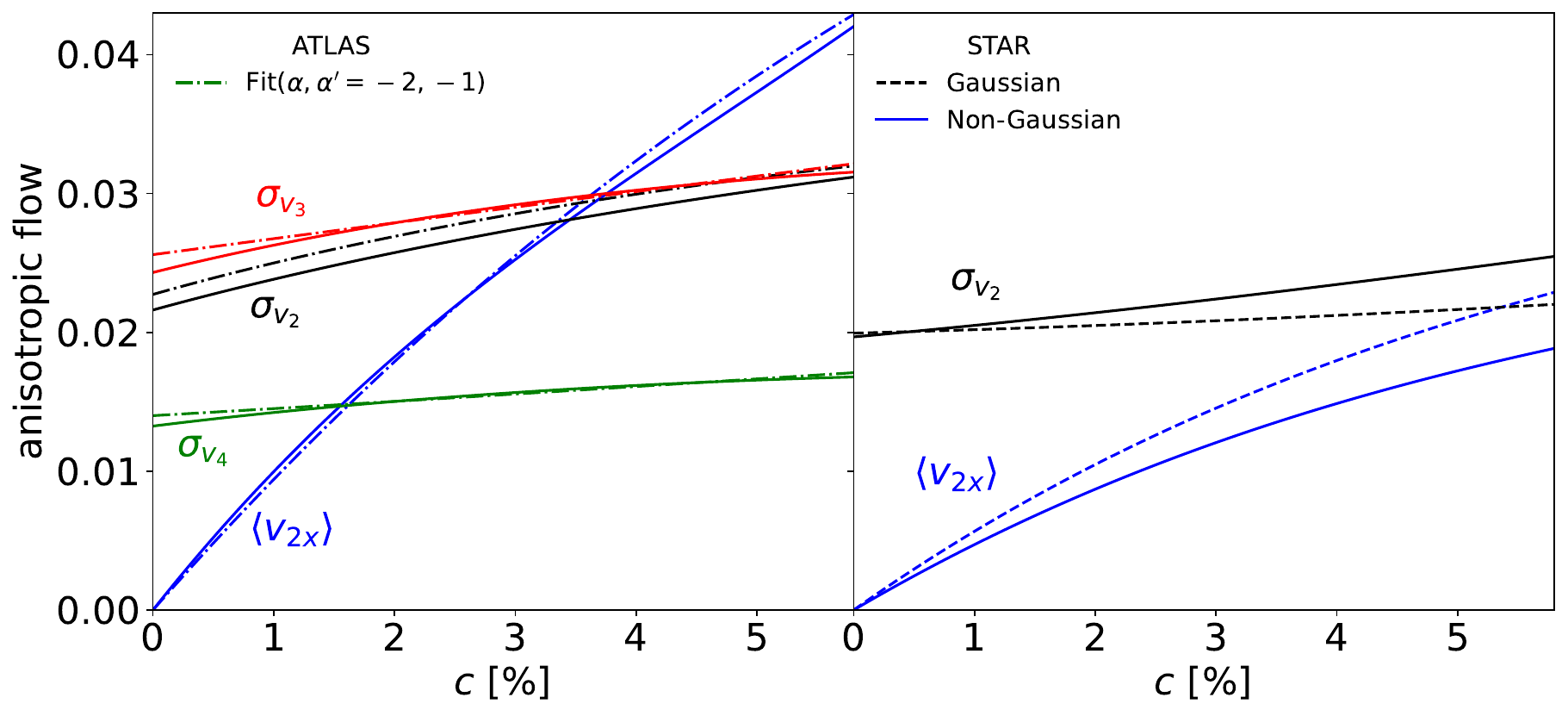} 
\end{center}
\caption{ 
Centrality dependence of the mean elliptic flow, and of the rms width of anisotropic flow fluctuations reconstructed by fitting ATLAS data in Fig.~\ref{fig:ATLAS} (left panel), and by fitting STAR data in Fig.~\ref{fig:STAR} (right panel). 
The different kinematic cuts ($0.5<p_t<5$~GeV/$c$ for ATLAS data, $0.2<p_t<2$~GeV/$c$ for STAR data) and, to a lesser extent, the different collision energies, explain why the STAR collaboration measure $v_2$ values that are smaller than those measured by the ATLAS collaboration. 
For ATLAS data, the full lines correspond to the reconstruction done as explained in this section. 
The dash-dotted lines display the modification induced by correlations between $v_n$ and multiplicity, discussed in Secs.~\ref{s:correlations} and \ref{s:improved}. 
For STAR data, the dashed lines correspond to the reconstruction done as explained in this section. 
The full lines display the modifications induced by taking into account the non-Gaussian corrections discussed in Sec.~\ref{s:nongaussian}. 
\label{fig:fixedb}
}
\end{figure*}    

The fit to data returns preferred values for the set of parameters $a_i$ and $A_i$ in  Eqs.~(\ref{polynom1}) and (\ref{polynom2}). 
Using these equations, we reconstruct the centrality dependence of the mean elliptic flow in the reaction plane, $\bar v_2(c)$, and of the standard deviation of flow fluctuations $\sigma_{v_n}(c)$, which are displayed in Fig.~\ref{fig:fixedb}.  

The magnitude of $\sigma_{v_n}(c=0)$ roughly matches that of $v_n\{2\}$ for the rightmost points in Figs.~\ref{fig:ATLAS} and \ref{fig:STAR}. 
This is intuitive as these points correspond essentially to collisions at zero centrality, where anisotropic flow is solely due to fluctuations.  
For ATLAS data, $\sigma_{v_3}$ and $\sigma_{v_2}$ are of comparable magnitude,\footnote{Note that the ATLAS analysis only includes particles with $p_t>0.5$~GeV/$c$. Since $v_3(p_t)$ increases with $p_t$ more steeply than $v_2(p_t)$, the $p_t$ cut enhances $v_3$ relative to $v_2$.} 
while $\sigma_{v_4}$ is smaller by approximately a factor two.

Still for ATLAS data, one observes that $\sigma_{v_n}$ increases as a function of centrality for all harmonics. 
This is natural since larger $c$ implies a smaller system size, hence larger fluctuations. 
When the initial density profile is the sum of $N$ independent sources, the variance of eccentricity fluctuations varies like $1/N$~\cite{Bhalerao:2006tp}  [see Eq.~(\ref{cumulantmatrix})]. 
Therefore, one typically expects that $\sigma_{v_n}^2(c)$ is inversely proportional to the mean multiplicity, $\bar N(c)$. 

The increase of fluctuations seen in Fig.~\ref{fig:fixedb} is in fact much faster than expected from this simple baseline. 
The relative increase of $\sigma_{v_n}^2(c)$ for central collisions is quantified by $d\ln\sigma_{v_n}^2/dc|_{c=0}=A_1/A_0$, where we have used Eq.~(\ref{polynom2}). 
The fit to ATLAS data returns values in the range $13<A_1/A_0<22$, depending on the harmonic.
This means that a $1\%$ increase in the centrality results in a $13\%$ to $22\%$ increase in the variance of flow fluctuations. 
This must be compared with the relative increase in  $1/\bar N(c)$, which is only $4\%$~\cite{Yousefnia:2021cup}. 
Thus, data indicates that flow fluctuations are suppressed in central collisions relative to the simple baseline of independent sources. 
As will be shown in Sec.~\ref{s:trento}, the same phenomenon is observed in simulations.

The increase of  $\sigma_{v_2}^2$ with centrality is  weaker for STAR data. 
However, our reconstruction is in this case less robust than that obtained for ATLAS data for the reasons outlined in Sec.~\ref{s:centrality}.

The mean elliptic flow in the reaction plane, $\bar v_2(c)$, overrides $\sigma_{v_2}(c)$ already at 4\% centrality for ATLAS data. 
In the case of STAR data, the crossing between $\bar v_2(c)$ and $\sigma_{v_2}(c)$ occurs for a larger value of $c$, which is however less precisely determined.

\subsection{Explaining the change of sign of $c_2\{4\}$}
\label{s:c24}

We now explain the origin of the change of sign of $nc_2\{4\}$ from negative to positive as a function of $N_{ch}$ or $E_T$, as shown  in Fig.~\ref{fig:ATLAS}.

Inserting the moments of Eq.~(\ref{gaussmoments}) into the definition of $c_n\{4\}$ in Eq.~(\ref{defcumulants}), one obtains: 
\begin{eqnarray}
  \label{cumul4}
  c_n\{4\} &=& 2\left(\langle \sigma_v^4\rangle-\langle\sigma_v^2\rangle^2\right)\cr
&&+4\left(\langle \sigma_v^2\bar v^2\rangle-\langle \sigma_v^2\rangle\langle\bar v^2\rangle\right) \cr
&&+  \langle\bar v^4\rangle-2\langle\bar v^2\rangle^2, 
\end{eqnarray}
where angular brackets denote an average over $c$ at fixed $N$. 
Thus $c_n\{4\}$ is decomposed as the sum of three terms. 
The first term involves fluctuations of $\sigma_v^2$, 
the second term the correlation between $\sigma_v^2$ and $\bar v^2$, 
while the third term is solely due to $\bar v$. 

For harmonics  $n=3$ and $4$, $\bar v$ vanishes, and only the first term involving the fluctuations of $\sigma_{v}^2$ remains. 
It is always positive, as can be seen in the red and green dashed lines of the corresponding panels of Fig.~\ref{fig:ATLAS}. 
The value of $nc_n\{4\}$ is of the same order of magnitude for both the harmonics, typically 1 or 2\%.  

For $n=2$, on the other hand, due to the nonzero $\bar v$, the first term is typically the smallest of the three. 
For the smallest values of $N_{ch}$ or $E_T$ (off-central collisions) in Fig.~\ref{fig:ATLAS}, the dominant contribution is the third term, so that $c_2\{4\}\simeq -\bar v^4$~\cite{Voloshin:2007pc} 
upon neglecting the variation of $\bar v$ with centrality.
Next comes the second term involving the covariance of $\sigma_v^2$ ad $\bar v^2$. 
Since both $\sigma_v^2$ and $\bar v^2$ increase with $c$, it is positive.
As $N_{ch}$ or $E_T$ increases (central collisions), it gradually overrides the third term, and explains the positive sign of $c_2\{4\}$ at large $N_{ch}$ or $E_T$. 

We now derive an approximate analytic estimate of $c_2\{4\}$ in this region. 
We neglect the first term of Eq.~(\ref{cumul4}). 
We evaluate the second term by expanding $\sigma_v^2(c)$ and $\bar v$ to first order in $c$: 
\begin{eqnarray}
\label{linearization}
\sigma^2_v(c)&\simeq &\sigma^2_v(0)+\left.\frac{d\sigma^2_v}{dc}\right|_{c=0} c\cr
\bar v(c)&\simeq &\left.\frac{d\bar v}{dc}\right|_{c=0} c. 
\end{eqnarray}
For the second line in Eq.~(\ref{cumul4}), one obtains
\begin{equation}
\label{2ndline}
\langle \sigma_v^2\bar v^2\rangle-\langle \sigma_v^2\rangle\langle\bar v^2\rangle\simeq \frac{d\sigma^2_v}{dc}\left(\frac{d\bar v}{dc}\right)^2 \left(\langle c^3\rangle-\langle c\rangle\langle c^2\rangle\right),
\end{equation}
where the derivatives are evaluated at $c=0$. 

This expression can be further simplified using the property that the distribution of $c$ is approximately Gaussian, Eq.~(\ref{gaussianc}). 
For values of $N$ small enough that the truncation of the Gaussian at $c=0$ has a negligible effect (blue lines in Fig.~\ref{fig:probc}), the mean centrality $\bar c(N)$ coincides with the experimentally defined centrality  $c_{\rm exp}$~\cite{Broniowski:2001ei}. 
Decomposing $c=c_{\rm exp}+\delta c$, and using $\langle \delta c\rangle=0$,  $\langle \delta c^2\rangle=\sigma_c^2$, $\langle \delta c^3\rangle=0$, one obtains: 
\begin{equation}
\label{2ndlinebis}
\langle c^3\rangle-\langle c\rangle\langle c^2\rangle=2\sigma_c^2c_{\rm exp}.
\end{equation}

We finally approximate the third line of Eq.~(\ref{cumul4}) by using the linear expansion of $\bar v(c)$ in Eq.~(\ref{linearization}) and neglecting the centrality fluctuation, that is, we replace $c$ with $c_{\rm exp}$: 
 \begin{equation}
\label{3rdline}
\langle\bar v^4\rangle-2\langle\bar v^2\rangle^2\simeq 
-\left(\frac{d\bar v}{dc}\right)^4 c_{\rm exp}^4. 
\end{equation}

Inserting Eqs.~(\ref{2ndline}), (\ref{2ndlinebis}) and (\ref{3rdline}) into Eq.~(\ref{cumul4}), we finally obtain: 
\begin{equation}
\label{approxcumul4}
c_2\{4\}\simeq -\left(\frac{d\bar v}{dc}\right)^4 c_{\rm exp}^4+  8\frac{d\sigma^2_v}{dc}\left(\frac{d\bar v}{dc}\right)^2 \sigma_c^2\, c_{\rm exp} .
\end{equation}
We recall that the derivatives in this expression are evaluated at $c=0$.
In this expression, only $c_{\rm exp}$ depends on the value of $N_{ch}$ or $E_T$. 
In addition, $\sigma_c$ depends on whether the centrality estimator is $N_{ch}$ or $E_T$.

The first term in the right-hand side of Eq.~(\ref{approxcumul4}) is the  negative contribution from the mean elliptic flow in the reaction plane~\cite{Borghini:2001vi}. 
The second term is positive and proportional to $\sigma_c^2$, which means that it comes from centrality fluctuations. 
Inspection of Eq.~(\ref{approxcumul4}) shows that the change of sign of $c_2\{4\}$ occurs at a centrality $c_{\rm exp}\propto (\sigma_c)^{2/3}$. 
The poorer the centrality resolution, the earlier  $c_2\{4\}$ becomes positive. 
This explains why a positive $v_2\{4\}$ signal is in general not observed at RHIC in the 0-5\% centrality window~\cite{STAR:2004jwm,STAR:2021mii}. It is only found for U+U collisions due to strong nuclear deformation effects~\cite{STAR:2015mki}.

Inspection of Eq.~(\ref{approxcumul4}) further shows that the value of $c_{\rm exp}$ where $c_2\{4\}$ reaches its maximum is smaller by a factor $2^{2/3}\simeq 1.6$ than the value at which it changes sign. 
The maximum value of $c_2\{4\}$ is proportional to $\sigma_c^{8/3}$. 
Using the values $\sigma_c\simeq 1.2\%$ for $N_{ch}$, $\sigma_c\simeq 0.85\%$ for $E_T$, one obtains that it is larger by a factor $\simeq 2.4$ for $N_{ch}$ than for  $E_T$, consistent with ATLAS data in Fig.~\ref{fig:ATLAS}. We provide, thus, a sound understanding of the detailed features that characterize the experimental measurements.  

Perhaps more importantly, the maximum value of $c_2\{4\}$ is directly proportional to $d\sigma_v^2/dc$, that is, to the increase of flow fluctuations with centrality. 
We have pointed out at the end of Sec.~\ref{s:centralitydependence}
 that this increase is faster than one would naively expect from an independent source model. 
We now see that a faster increase also results in a larger   positive  $nc_2\{4\}$. A similar reasoning could be applied to the cumulant of order six, $c_2\{6\}$, which we leave for future work. 
Note that according to our numerical calculation in Fig.~\ref{fig:ATLAS}, it changes sign twice. It would be insightful to improve the precision of this measurement with future LHC Run3 Pb+Pb data, whose analysis is underway.

\subsection{Non-Gaussian corrections}
\label{s:nongaussian}

The picture of Gaussian flow fluctuations presented in Sec.~\ref{s:gaussian} is motivated by the central limit theorem, applied to fluctuations in the early stages of the collision. 
The collision involves nonetheless only a few hundred nucleons, such that finite-size corrections are expected in the form of higher-order cumulants, the first two being the skewness and the kurtosis.\footnote{The term ``kurtosis'' sometimes refers to the central moment of order four, while the cumulant is called the ``excess kurtosis''. Throughout this paper, the kurtosis denotes the cumulant.}
There is no general theorem for the signs of skewness and kurtosis of a random variable, but there are systematic trends depending on the support of the underlying distribution. 
For a positive variable, such as the particle multiplicity~\cite{Rogly:2018ddx} or the transverse momentum per particle~\cite{Samanta:2023kfk,Giacalone:2020lbm}, the skewness and kurtosis are usually positive.\footnote{A typical example is the Poisson distributions, whose all cumulants are equal.}
On the other hand, for a variable which lies on the unit disk, such as the complex flow $V_n$ or the initial anisotropy $\varepsilon_n$ (Appendix \ref{s:cumulants}),  skewness~\cite{Yan:2014afa,Giacalone:2016eyu,Mehrabpour:2018kjs,Cirkovic:2019yvs} and kurtosis~\cite{Yan:2013laa} are usually negative. 

When anisotropic flow is only due to fluctuations, there is no skewness by symmetry, but one expects a negative kurtosis.  
This was first observed by the ALICE collaboration for $v_3$ in Pb+Pb collisions~\cite{ALICE:2011ab}, in the form of a positive $v_3\{4\}$, which was then unexpected. 
The negative kurtosis of $v_2$ and $v_3$ was subsequently observed in p+Pb collisions~\cite{ATLAS:2013jmi,CMS:2013jlh,ALICE:2014dwt,CMS:2015yux,ATLAS:2017rtr,CMS:2017glf,CMS:2019wiy}, where the smaller system size implies even larger non-Gaussianities. In the case of elliptic flow in nucleus-nucleus collisions, higher-order cumulants are generated by elliptic flow in the reaction plane, as shown in Sec.~\ref{s:gaussian}. 
They do not directly reflect non-Gaussian fluctuations, whose signatures are somewhat elusive. 
The skewness of the elliptic flow vector distribution along the direction of the reaction plane can be inferred from the small splitting between $v_2\{4\}$ and $v_2\{6\}$ in mid-central Pb+Pb collisions~\cite{CMS:2017glf,ALICE:2018rtz}. 
Similarly, the kurtosis can be extracted, albeit with complications~\cite{Bhalerao:2018anl,CMS:2023bvg}. 

The leading non-Gaussian corrections relevant for central nucleus-nucleus collisions of interest in the present analysis are discussed in Appendix~\ref{s:cumulants}. 
We discuss cumulants of the initial anisotropies, $\varepsilon_n$, but the formalism is identical for cumulants of anisotropic flow, upon replacing $\varepsilon\equiv\varepsilon_n$ by $v$ everywhere. 
We parametrize the skewness and the kurtosis via two parameters ${\tt Skw}_v$ and ${\tt Krt}_v$. 
They are normalized in such a way that they depend weakly on centrality. 
They are {\it intensive\/} quantities, in the same way as the intensive skewness of $p_t$ fluctuations~\cite{Giacalone:2020lbm,ALICE:2023tej}. 

We do not write explicitly the corrected form of the Gaussian distribution in Eq.~(\ref{gaussianv}), because we are only interested in its moments. 
The latter are simply expressed in terms of cumulants using the generating function, as explained in detail in Appendix~\ref{s:cumulants}. 
The skewness and kurtosis appear as additional contributions to the moments of order 4 and higher. This amounts to replacing Eqs.~(\ref{gaussmoments}) with (see Appendix~\ref{s:cumulants}): 
\begin{widetext}
\begin{eqnarray}
  \label{nongaussmoments}
  \langle |V|^2|c\rangle&=& \sigma_v^2+\bar v^2\cr
  \langle |V|^4|c\rangle&=&\left(2+\sigma_v^2{\tt Krt}_v \right)\sigma_v^4+4\left(1+\sigma_v^2{\tt Skw}_v\right) \sigma_v^2\bar v^2+
  \bar v^4  \cr
  \langle |V|^6|c\rangle&=&\left(6+9 \sigma_v^2{\tt Krt}_v\right)\sigma_v^6+9 \left(2+\sigma_v^2({\tt Krt}_v+4{\tt Skw}_v)\right)
  \sigma_v^4\bar v^2+9\left(1+2\sigma_v^2{\tt Skw}_v \right)\sigma_v^2\bar v^4+
  \bar v^6.   
\end{eqnarray}
\end{widetext}
For $v_3$ and $v_4$, the mean anisotropy $\bar v$ vanishes, and the moments only depend on $\sigma_v$ and on the kurtosis. 

\begin{table}
\begin{tabular}{|c|c|c|c|}
\hline
&$v_2$&$v_3$&$v_4$\cr
\hline
$a_1$&$1.10\pm 0.01$&&\cr
&$\mathit{0.96\pm 0.04}$&&\cr
$a_2$&$-10.7\pm 0.3$&&\cr
&$\mathit{-4.9\pm 1.2}$&&\cr
$a_3$&$75\pm 3$&&\cr
&$\mathit{11\pm 12}$&&\cr
$10^4 A_0$&$4.67\pm 0.01$ &$5.91\pm 0.01$&$1.754\pm 0.005$\cr
&$\mathit{5.17\pm 0.03}$&$\mathit{6.55\pm 0.01}$&$\mathit{1.96\pm 0.02}$\cr
$10^3 A_1$
&$10.3\pm 0.3$
&$10.6\pm 0.1$
&$2.86\pm 0.06$\cr
&$\mathit{11.2\pm 0.7}$&$\mathit{5.9\pm 0.2}$&$\mathit{1.4\pm 0.1}$\cr
$10^2 A_2$
&$-2.8\pm 0.4$
&$-6.2\pm 0.2$
&$-1.8\pm 0.1$\cr
&$\mathit{-4.3\pm 0.9}$&$\mathit{1.0\pm 0.3}$&$\mathit{0.4\pm 0.2}$\cr
${\tt Skw}_v$
&$-36\pm 5$
&
&\cr
&$\mathit{25 \pm 8}$&&\cr
${\tt Krt}_v$
&$-18\pm 5$
&$-29\pm 1$
&$-243\pm 10$\cr
&$\mathit{39\pm 17}$&$\mathit{-12\pm 2}$&$\mathit{-197\pm 17}$\cr
$\chi^2/$dof
&$2.9$
&$1.7$
&$1.4$\cr
&$\mathit{4.2}$&$\mathit{11.7}$&$\mathit{4.3}$\cr
\hline
\end{tabular}
\caption{\label{fitparameters} 
Values of the fit parameters defined by Eqs.~(\ref{polynom1}), (\ref{polynom2}) and (\ref{nongaussmoments}), and of the chi square per degree of freedom, for the ``combined'' fit to ATLAS data in Fig.~\ref{fig:ATLAS}. 
Two sets of parameters are provided. 
The first set, in roman characters, corresponds to the procedure described in Sec.~\ref{s:nongaussian}, where the (non-Gaussian) probability distribution of anisotropic flow is assumed to depend only on centrality (full lines in Fig.~\ref{fig:fixedb})  
The second set, in italics, assumes a specific correlation between anisotropic flow and the centrality estimator $N$, which is discussed in Sec.~\ref{s:correlations} (dash-dotted lines in Fig.~\ref{fig:fixedb}). 
}
\end{table}

Taking into account these non-Gaussian corrections amounts to adding one more fit parameter, ${\tt Krt}_v$, for both $v_3$ and $v_4$, and two more fit parameters,  ${\tt Skw}_v$ and ${\tt Krt}_v$, for $v_2$. 
The resulting fits to ATLAS data are displayed as solid lines in Fig.~\ref{fig:ATLAS}, and the values of fit parameters are listed in Table~\ref{fitparameters}. 
The difference with respect to the Gaussian fit is significant only for the cumulants of order four or higher. 
The negative kurtosis significantly improves agreement with data for $nc_3\{4\}$ and $nc_4\{4\}$, as expected.
This is reflected by the chi square of the fit, which is reduced from about $20$ to less than $2$. 
Note, however, that the fit does not capture the variation pattern of $nc_4\{4\}$ when the centrality estimator is $N_{ch}$, which is less regular than when the centrality estimator is $E_T$. 
For elliptic flow, the improvement brought by the non-Gaussian terms is marginal. 
However, it is interesting to notice that the fit returns negative values of ${\tt Skw}_v$ and ${\tt Krt}_v$, which is also expected, albeit with large error bars. 
In the case of STAR data, the improvement brought by the non-Gaussian correction is not visible by eye.
The fit returns a positive value of the skewness, and a large negative value of the kurtosis. 

The values of the fit parameters entering Eqs.~(\ref{polynom1}) and (\ref{polynom2}) differ depending on whether or not non-Gaussian corrections are added. 
For ATLAS data, the change is small enough that it would be barely visible on Fig.~\ref{fig:fixedb}, where we actually display the results obtained {\it with\/} the non-Gaussian terms included. 
For STAR data, both sets of results are shown, and the difference is significant. 
It shows that the reconstruction has large uncertainties, as anticipated from the discussion in Sec.~\ref{s:centrality}.

We now interpret the values of the skewness and kurtosis returned by our fit to ATLAS data in a simple hydrodynamic picture, where anisotropic flow results from a linear hydrodynamic response to an initial anisotropy: 
\begin{equation}
\label{linearresponse}
V_n=\kappa_n \varepsilon_n,
\end{equation}
where $V_n$ is the complex flow, $\varepsilon_n$ the complex eccentricity, and $\kappa_n$ a (real) hydrodynamic response coefficient, which depends on $n$~\cite{Teaney:2012ke}, usually smaller than unity. This is a good approximation for $n=2$ and $n=3$ both in ideal and viscous hydrodynamics~\cite{Niemi:2012aj}.  
For $n=4$, there are corrections due to a nonlinear coupling with elliptic flow~\cite{Gardim:2011xv,Teaney:2012ke}, but they are negligible for central Pb+Pb collisions, on which we focus now. 
Within this linear response scenario, Eq.~(\ref{defcmn}) gives
\begin{equation}
\label{cumulantslinear}
c_{mp}(v_n)=\kappa_n^{m+p}c_{mp}(\varepsilon_n), 
\end{equation}
where we denote by $c_{mp}(v_n)$ the cumulants of the (two-dimensional distribution) of $V_n$ and by $c_{mp}(\varepsilon_n)$ those of the distribution of $\varepsilon_n$. 
Using Eq.~(\ref{defintensive}), where $\bar v_n=\kappa_n\bar\varepsilon_n$ and $\sigma_{v_n}=\kappa_n\sigma_{\varepsilon_n}$, one then obtains: 
\begin{eqnarray}
\label{linearnongaussian}
{\tt Skw}_{v_n}&=&\frac{1}{\kappa_n^2} {\tt Skw}_{\varepsilon_n}\cr
{\tt Krt}_{v_n}&=&\frac{1}{\kappa_n^2} {\tt Krt}_{\varepsilon_n},
\end{eqnarray}
which means that the intensive skewness and kurtosis of flow are amplified by a factor $1/\kappa_n^2$ relative to those of the initial eccentricity. 

Now, we shall see in Sec.~\ref{s:nongaussianT} that initial-state models return values of 
\begin{equation}
\label{EllipticPower}
 {\tt Skw}_{\varepsilon_n}\approx {\tt Krt}_{\varepsilon_n}\approx -2. 
\end{equation}
In the case of the Elliptic-Power distribution~\cite{Yan:2014afa}, which is a generic parametrization of the distribution of $\varepsilon_n$,  the cumulants can be calculated analytically~\cite{Bhalerao:2018anl}, and one obtains precisely  ${\tt Skw}_{\varepsilon_n}={\tt Krt}_{\varepsilon_n}= -2$ in the limit $\bar\varepsilon\ll 1$ and $N\gg 1$, where $N$ is the number of sources.  
Putting together Eqs.~(\ref{linearnongaussian}) and (\ref{EllipticPower}), one then obtains the linear response coefficients from the kurtosis as: 
\begin{equation}
\label{kappafromnongauss}
\kappa_n\approx\sqrt{\frac{-2}{{\tt Krt}_{v_n}}},
\end{equation}
and a similar equation for the skewness. 
Using the values of ${\tt Krt}_{v_n}$ in Table~\ref{fitparameters}, we obtain
\begin{eqnarray}
\label{kappafromfit}
0.26&<&\kappa_3<0.41 \cr 
0.09&<&\kappa_4<0.11.
\end{eqnarray}
The error bars on ${\tt Skw}_{v_2}$ and ${\tt Krt}_{v_2}$ are too large to obtain a meaningful estimate of $\kappa_2$. 
The estimates in Eq.~(\ref{kappafromfit}) are in the ballpark of hydrodynamic results~\cite{Teaney:2012ke}, if one properly takes into account the kinematic cut $p_t>0.5$~GeV/$c$ of the ATLAS analysis, which preferentially selects particles with higher $v_n$. 
It is interesting to note that the uncertainty is smallest for the highest harmonic $\kappa_4$, which is also the most sensitive to viscosities in the hydrodynamic picture. 
Hence, the non-Gaussianity of the quadrangular flow distribution~\cite{Giacalone:2016mdr} gives rather direct access to the hydrodynamic response, and can potentially help tighten the experimental constraints on transport coefficients of QCD from Bayesian analyses~\cite{Paquet:2023rfd}. 
Before drawing definite conclusions, however, hydrodynamic simulations at fixed impact parameter should be carried out in order to test whether higher-order cumulants of anisotropic flow are accurately determined by linear response to the eccentricity, as assumed in Eq.~(\ref{cumulantslinear}).

\section{Models of initial fluctuations} 
\label{s:trento}

\subsection{Motivation}
\label{s:motivation}

We assess the accuracy of the reconstruction carried out in Sec.~\ref{s:data} using model simulations where the impact parameter is known in each event. 
Ideally, these model simulations should be event-by-event hydrodynamic  simulations~\cite{Schenke:2020mbo}, in which one calculates the particle multiplicity as well as the anisotropic flow $v_n$, for a large number of simulated events. 
One can then test the reconstruction by carrying out two sets of simulations. 
First, one can generate events randomly, mimicking minimum-bias events in an actual experiment. 
These events can then be analyzed in the same way as in an experiment, by binning them according to the particle multiplicity and evaluating the cumulants [Eq. (\ref{defcumulants})] in each bin. 
Second, one can also generate events at fixed impact parameters, which cannot be done experimentally. 
With this second set of simulations, one can study the fluctuations of multiplicity and anisotropic flow at a fixed impact parameter to test the reconstructed results.   

Full event-by-event hydrodynamic simulations are numerically heavy, and we choose to test our assumptions using a simpler framework. 
First, as done in the previous section, we consider that final-state anisotropic flow coefficients, $V_n$, originate from a linear response to the initial-state anisotropies, $\varepsilon_n$.
Furthermore, we consider that the charged multiplicity of an event, such as $N_{\rm ch}$, is largely determined by the initial entropy, obtained by integrating the entropy density deposited in the collision process at midrapidity over the transverse coordinates. 
Therefore, we choose to simulate only the initial conditions of the heavy-ion collision.
Such a simulation is not as realistic as a full hydrodynamic calculation, but we hope that it captures the salient features, in particular the effect of the geometry, as mediated by impact parameter, and the event-by-event fluctuations.

\subsection{Modeling initial conditions}
\label{s:trentosetup}

We use the \trento{} model, which is a Monte Carlo generator of initial  density profiles (entropy profiles, in our case) in high-energy nuclear collisions~\cite{Moreland:2014oya}, routinely employed in state-of-the-art simulations~\cite{JETSCAPE:2020mzn,Giacalone:2023cet}.\footnote{Note that recent works use  \trento{} as an initial condition for the energy density, rather than the entropy density.}

The model first samples the positions of nucleons within each nucleus. 
Then, it defines thickness functions $T_{A,B}({\bf x_\perp})$ for each nucleus ($A$ and $B$, respectively) according to the positions of these nucleons, where ${\bf x_\perp}$ denotes the transverse coordinate. 
We assume that the entropy density $s({\bf x_\perp})$ is proportional to the geometric mean of these thickness functions,  $s\propto \sqrt{T_A T_B}$, which corresponds to the option $p=0$ in the code. 
The contribution of each nucleon to the thickness function is allowed to fluctuate randomly, and these fluctuations are controlled by a parameter $k$.\footnote{Additional sources of fluctuations at the nucleon level can be introduced by adding a fluctuating sub-nucleonic structure to the colliding nucleons \cite{Moreland:2018gsh}. For the purpose of this study, it is enough to consider smooth structureless nucleons, modeled as Gaussian profiles. The corresponding nucleon size is $w=0.5$ fm.}
 We choose the value $k=2$, for which the relative fluctuations of the entropy at $b=0$ have the same magnitude as those of the charged multiplicity in ATLAS data (specifically, $\sigma_{N_{ch}}(0)/\overline{N_{ch}}(0)\simeq 4\%$~\cite{Yousefnia:2021cup}). 

We run the \trento{} model for Pb+Pb collisions at $\sqrt{s_{NN}}=5.02$~TeV. 
We first generate $4 \times 10^8$ minimum-bias events. 
In each event, we evaluate the total entropy, $S$, and the initial eccentricities $|\varepsilon_n|$, for $n=2,3,4$.  
We assume that the charged multiplicity $N_{ch}$ of the event is proportional to $S$. 
We fix the proportionality factor by running  \trento{} simulations at $b=0$ (see below) and evaluating the average value of $S$ for these collisions, $\bar S (b=0)$. 
Now, the average value of $N_{ch}$ for $b=0$ can be reconstructed from the histogram of $N_{ch}$ displayed in the upper left panel of Fig.~\ref{fig:ATLAS}, as explained in Sec.~\ref{s:centrality}, and one obtains the value $\overline{N_{ch}}(b=0)\simeq 3104$~\cite{Yousefnia:2021cup}. 
We choose the proportionality factor as $\overline{N_{\rm ch}}(b=0)/\bar S (b=0)$,  and rescale the minimum-bias distribution of $S$ values accordingly. 

\begin{figure}[t]
\begin{center}
\includegraphics[width=\linewidth]{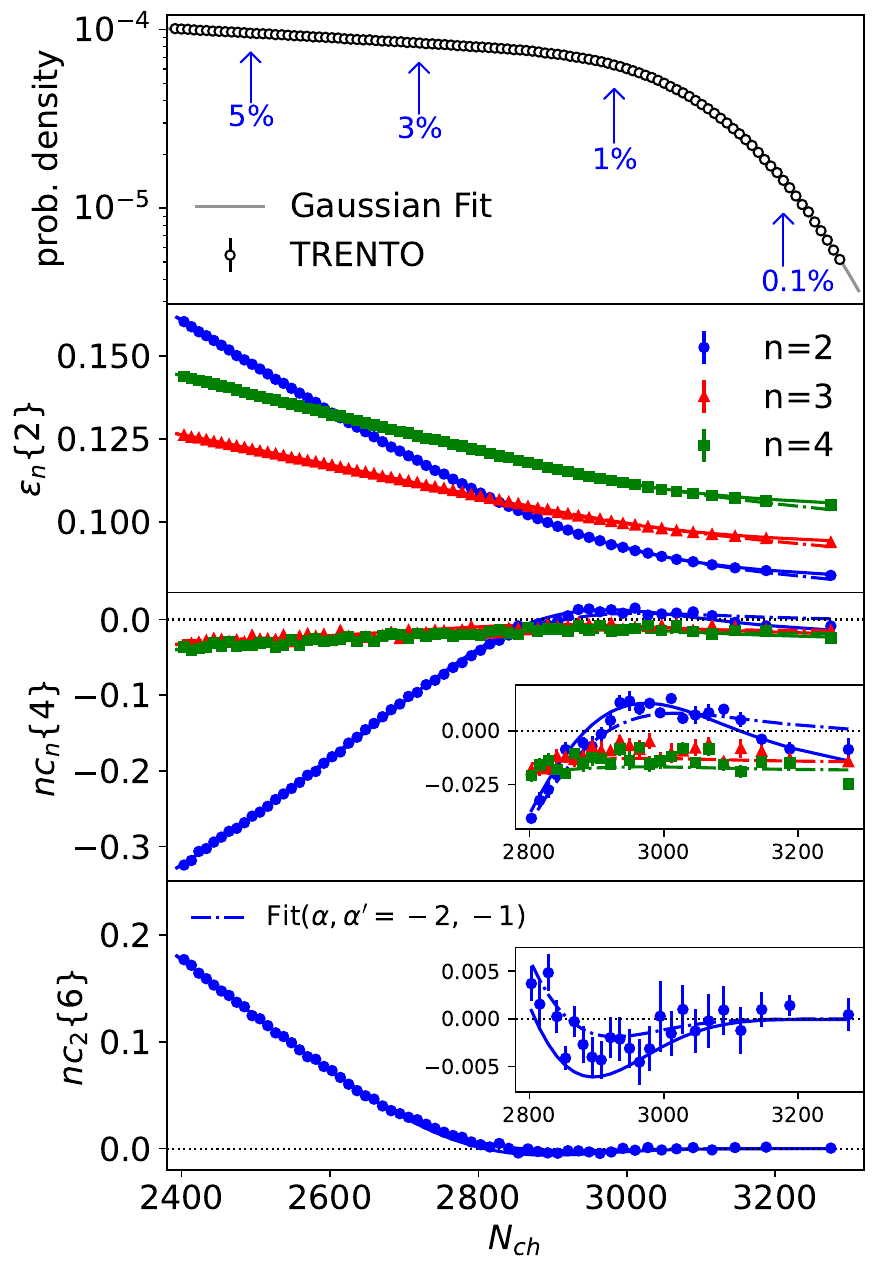} 
\end{center}
\caption{ Top: distributions of the charged multiplicity $N_{ch}$ obtained in the \trento{} simulation from a rescaling of the initial-state entropy, as explained in the text. The solid gray line is a fit using Eqs.~(\ref{gaussian}) and (\ref{fitpn}), as in the case of ATLAS and STAR data in Figs.~\ref{fig:ATLAS} and \ref{fig:STAR}. 
Lower panels display the cumulants of eccentricity fluctuations ($\varepsilon_n$, $n=2$, 3, 4) as a function of $N_{ch}$, defined by Eqs.~(\ref{defcumulants}) and (\ref{defncumulants}) with $v_n \to \varepsilon_n$.
Lines are fits using Eqs.~(\ref{nongaussmoments}). 
Full lines are obtained by assuming that the distribution of $\varepsilon_n$ only depends on centrality, as in Sec.~\ref{s:data}. 
Dash-dotted lines are obtained by modeling the correlation between $\varepsilon_n$ and $N_{\rm ch}$ using Eqs.~(\ref{alphaexp}) and (\ref{alphapexp}), with $\alpha=-2$ and $\alpha'_2=\alpha'_3=-1$ (see text). 
}
\label{fig:cumulantsT}
\end{figure} 

Figure~\ref{fig:cumulantsT} displays the distribution of $N_{ch}$ from this model calculation (upper panel), as well as the cumulants of the eccentricity distributions (middle and lower panels), obtained by replacing $V_n$ with $\varepsilon_n$ in Eqs.~(\ref{defcumulants}) and (\ref{defncumulants}). 
The similarity with Fig.~\ref{fig:ATLAS} is obvious. 
The magnitude of $\varepsilon_n\{2\}$ is significantly larger than that of $v_n\{2\}$, as expected in hydrodynamic models, where the response coefficient $\kappa_n=v_n\{2\}/\varepsilon_n\{2\}$ is smaller than unity. 
The higher-order cumulants of $\varepsilon_2$, $nc_2\{4\}$ and $nc_2\{6\}$, change sign like the cumulants of $v_2$ in Fig.~\ref{fig:ATLAS}. 
If $v_n$ was strictly proportional to $\varepsilon_n$, then the proportionality factor would cancel between the numerator and the denominator of Eq.~(\ref{defncumulants}), and $nc_n\{4\}$ and $nc_n\{6\}$ would be identical for $v_n$ and $\varepsilon_n$. 
One sees that $nc_2\{4\}$ reaches positive values significantly larger for $v_2$ in Fig.~\ref{fig:ATLAS} than for $\varepsilon_2$ in Fig.~\ref{fig:cumulantsT}. 
Note also that $nc_2\{4\}$ becomes negative again for the largest values of $N_{ch}$. 
This second change of sign, which is not seen in data (where it could be hidden by error bars), arises from the non-Gaussianity of $\varepsilon_n$ fluctuations. 
As will be discussed in detail in Sec.~\ref{s:nongaussianT}, these fluctuations have negative kurtosis at fixed impact parameter, and the highest values of $N_{ch}$ correspond to collisions at zero impact parameter. 

The differences between model and data for $nc_2\{4\}$ may be due to the fact that our model of initial conditions is not realistic.  
They might be also explained by deviations from the linear response. 
Recent calculations~\cite{Giacalone:2024ixe} have shown that in the case of ultracentral Pb+O collisions, $nc_2\{4\}$ is negative for $\varepsilon_2$ fluctuations but positive for $v_2$ fluctuations, meaning that the hydrodynamic response  indeed increases $nc_2\{4\}$. 
We do not investigate the origin of this interesting difference, as our goal is simply to test the reconstruction procedure on mock data.

\subsection{Reconstruction of the centrality dependence}
\label{s:firstreconstruction}

We then analyze the results of these simulations exactly as we analyze the ATLAS data. 
First, we fit the distribution of $N_{ch}$  as in  Sec.~\ref{s:centrality}, so as to reconstruct the distribution of centrality at fixed $N_{ch}$, $P(c|N_{ch})$. 
Next, we fit the results on cumulants using  Eqs.~(\ref{nongaussmoments}), where $v$ is replaced by $\varepsilon$ everywhere. 
The mean eccentricity $\bar\varepsilon_2(c)$ and the variance of fluctuations $\sigma^2_{\varepsilon_n}(c)$ are parametrized as polynomials as in Sec.~\ref{s:data}. 
The resulting fits, displayed as full lines in Fig.~\ref{fig:cumulantsT}, are excellent, as in the case of ATLAS data. 

\begin{figure*}[t]
\begin{center}
\includegraphics[width=.6\linewidth]{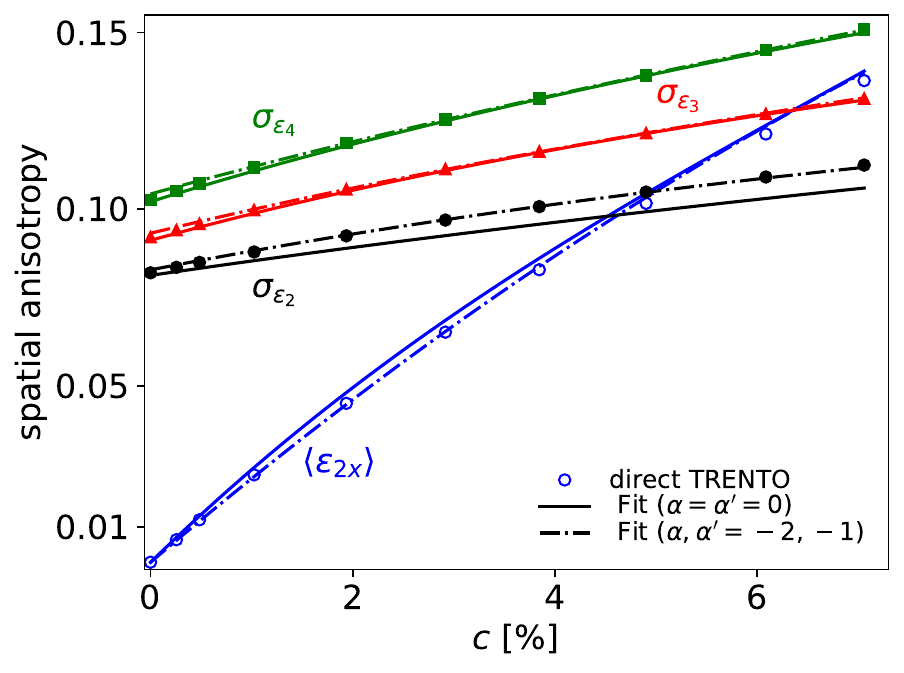} 
\end{center}
\caption{
Centrality dependence of the
mean eccentricity $\bar\varepsilon_2$, and of the rms width of eccentricity fluctuations, $\sigma_{\varepsilon_n}$, in our \trento{}  simulations. 
Symbols are values calculated directly by running the model for fixed centrality. 
Lines are obtained by fitting the mock data of Fig.~\ref{fig:cumulantsT}, either with (dash-dotted lines) or without (full lines) taking into account the correlation between eccentricity and multiplicity at fixed centrality (see text).}
\label{fig:fixedbT}
\end{figure*} 

The full lines in Fig.~\ref{fig:fixedbT} display the functions $\bar\varepsilon_2(c)$ and $\sigma_{\varepsilon_n}(c)$ for $n=2,3,4$, as returned by the fit. 
In order to test the accuracy of this reconstruction, we run the \trento{} model at fixed impact parameter. 
We generate  $4\times 10^6$ events for each value of $c$.\footnote{The \trento{} calculation returns $\sigma_{\rm inel}=785$ fm$^2$, in good agreement with the ALICE result \cite{ALICE:2022xir}.}  
The resulting values of $\bar\varepsilon_2(c)$ and $\sigma_{\varepsilon_n}(c)$ are displayed as symbols in Fig.~\ref{fig:fixedbT}. 
One sees that $\sigma_{\varepsilon_3}(c)$ and $\sigma_{\varepsilon_4}(c)$ are accurately reconstructed. 
But the reconstruction is not as precise in the second harmonic: 
Specifically, it slightly overestimates  $\bar\varepsilon_2(c)$ and underestimates  $\sigma_{\varepsilon_2}(c)$.

\subsection{Correlations between anisotropy and multiplicity}
\label{s:correlations}

The origin of this slight discrepancy can be traced back to the existence of a correlation between the initial anisotropy, $\varepsilon_n$, and the entropy, $S$, at fixed centrality, which we have not considered in our calculations. 
In other terms, in writing Eq.~(\ref{particular}) we have implicitly assumed that the multiplicity, $N$, and $V_n$ are independent variables at fixed $c$. 
Releasing this assumption, we generalize Eq.~(\ref{particular}) for the moments of the flow fluctuations:
\begin{equation}
\label{particular2}
  \langle |V|^k|N\rangle=\int \langle |V|^k|c,N\rangle  P(c|N)dc, 
\end{equation}
meaning the average of $|V|^k$ at fixed $c$ does in general depend on $N$. 
A similar equation holds if one replaces anisotropic flow, $V$, by the initial anisotropy, $\varepsilon$.

Since the distribution of $\varepsilon_n=(\varepsilon_{nx},\varepsilon_{ny})$ at fixed $c$, $P(\varepsilon_n|c)$, is Gaussian up to small corrections, which will be discussed below, it suffices to specify how the parameters of the Gaussian in Eq.~(\ref{gaussianv}) are correlated with $N$ (which in this case represents the initial-state entropy, $S$) at fixed $c$, for instance, the reaction plane eccentricity, $\bar\varepsilon_2$.
A simple and yet effective way to parametrize its correlation with $N$ is by means of a power law dependence, with an exponent $\alpha$:
\begin{equation}
  \label{alphaexp}
\bar\varepsilon_2(c,N)\equiv
\langle\varepsilon_{2x}|c,N\rangle
=\bar\varepsilon_2(c)\left(\frac{N}{\bar N(c)}\right)^{\alpha(c)}. 
\end{equation}
In practice, at fixed $c$ the fluctuations of $N$ around the average $\bar N(c)$ are small enough that one can linearize Eq.~(\ref{alphaexp}): 
\begin{equation}
  \label{alphaexplin}
\bar\varepsilon_2(c,N)\simeq \bar\varepsilon_2(c)\left(1+\alpha(c) \frac{N-\bar N (c)}{\bar N (c)}\right).  
\end{equation}
Averaging over $N$, the second term vanishes, and the average value of  $\bar\varepsilon_2(c,N)$ is $\bar\varepsilon_2(c)$ as it should. 
We now relate $\alpha(c)$ to the linear correlation between $\varepsilon_{2x}$ and $N$, which is the standard measure of a correlation. 
We multiply Eq.~(\ref{alphaexplin}) by $N$ and average over events at fixed $c$:  
\begin{equation}
\label{lincorr}
 \langle \varepsilon_{2x} N\rangle -\bar \varepsilon_{2}\bar N=\alpha 
 \bar\varepsilon_{2}\frac{\sigma_N^2}{\bar N}, 
\end{equation}
where the dependence on $c$ is implicit. 
We use this equation to evaluate $\alpha(c)$ numerically, by running simulations at fixed $c$. 

The exponent $\alpha$ is related to the traditional Pearson correlation, $\rho$, computed between $\varepsilon_{2x}$ and $N$ in the following way: 
\begin{equation}
\label{Pearson}
 \rho \equiv \frac{\langle \varepsilon_{2x} N\rangle - \bar\varepsilon_2\bar N}{\sigma_{2x} \, \sigma_{N}} = \alpha\frac{\sigma_N}{\bar N}\frac{\bar\varepsilon_2}{\sigma_{2x}},
\end{equation}
where $\sigma_{2x}\equiv\sqrt{\langle\varepsilon_{2x}^2\rangle-\bar\varepsilon_2}$ denotes the standard deviation of the reaction-plane eccentricity, $\varepsilon_{2x}$. 
In the limit $c\to 0$, azimuthal symmetry implies that $\rho$ vanishes, while $\alpha$ converges to a finite value. 
Therefore, characterizing this correlation via an effective exponent $\alpha$ as in Eq.~(\ref{lincorr}), rather than by the Pearson coefficient, seems more convenient for ultracentral collisions, as we expect its variation with the centrality to be negligible. Indeed, variables associated to fluctuations are expected to approach the ultra-central limit in a smooth, non-singular way.

A similar discussion applies to the correlation between $\sigma_{\varepsilon_n}^2$ and $N$, which we parameterize by an equation similar to (\ref{alphaexp}):  
\begin{equation}
\label{alphapexp}
 \sigma_{\varepsilon_n}^2(c,N)
 = \sigma_{\varepsilon_n}^2(c)\left(\frac{N}{\bar N(c)}\right)^{\alpha'_n(c)},
\end{equation}
with a new exponent $\alpha'_n(c)$. 
The value of $\alpha'_n$ is calculated using an equation similar to Eq.~(\ref{lincorr}), up to the following modifications. 
The left-hand side of Eq.~(\ref{lincorr}) is the connected correlation, $\langle \varepsilon_{2x}N\rangle_c$, which one should replace with the connected correlation $\langle (\varepsilon_{nx}^2+\varepsilon_{ny}^2)N\rangle_c$, obtained by subtracting lower-order correlations order by order (the explicit expression is derived in Appendix~\ref{s:cumulants}, Eq.~(\ref{c111})). 
In the right-hand side of Eq.~(\ref{lincorr}), one should simply replace $\bar\varepsilon_2(c)$ with $\sigma^2_{\varepsilon_n}(c)$.

\begin{figure}[t]
\begin{center}
\includegraphics[width=\linewidth]{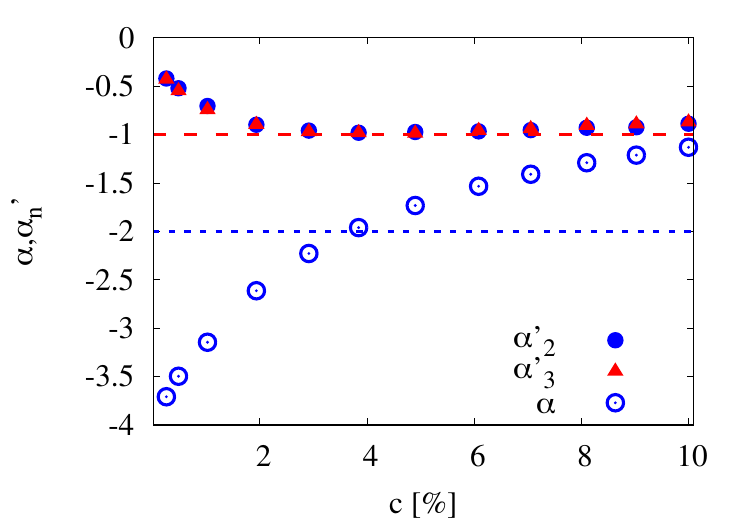} 
\end{center}
\caption{Centrality dependence of the exponents $\alpha$ [Eq.~(\ref{alphaexp})] and $\alpha'_n$, with $n=2,3$ [Eq.~(\ref{alphapexp})] in \trento{} simulations at fixed impact parameter. 
The symbols are direct numerical evaluations, while the lines indicate the values which we have used in order to evaluate the impact of these coefficients on our fits and the reconstruction procedure (see text). 
}
\label{fig:correlations}
\end{figure}
The values of $\alpha$ and $\alpha'_n$ obtained from  \trento{} model simulations at fixed centrality are displayed in Fig.~\ref{fig:correlations}. 
They are nonzero in magnitude and negative. 
Equations~(\ref{alphaexp}) and (\ref{alphapexp}) then imply that, 
for a given impact parameter, collisions producing a larger multiplicity have on average a smaller eccentricity in the reaction plane, as well as smaller eccentricity fluctuations. 
We have checked, by varying the parameters of the \trento{} model, that this is a generic feature. 
In particular, larger values of the fluctuation parameter $k$ (corresponding to lower event-by-event entropy density fluctuations) result in even more negative values of $\alpha$ and $\alpha'_n$. 
One sees also that the value of $\alpha'_n$ is almost identical for $n=2$ and $n=3$. 
Surprisingly, we discover that the exponents $\alpha$ and $\alpha'_n$ display a significant variation with $c$ in the $0-10\%$ window.\footnote{By contrast, they are almost constant for semi-central collisions covering the $10-40\%$ range.}
We do not understand the precise the origin of this variation. 
It implies that these variables approach the ultra-central limit in a rather singular manner. 
We suspect that it is related to the selection of participant nucleons~\cite{Miller:2007ri} that is the starting point of the \trento{} calculation. 
In particular, fluctuations in the number of participants should decrease drastically in the limit $c\to 0$, which may drive the singular behavior of other quantities. 
We leave the investigation of this phenomenon to a future study, as here we are only interested in knowing whether the inclusion of non-zero $\alpha$ and $\alpha'_n$ values improves the reconstruction of the eccentricities at fixed impact parameter, which we study now.

\subsection{Assessing the accuracy of the reconstruction}
\label{s:improved}
  
The values of $\alpha$ and $\alpha'_n$ can be calculated in simulations, but they are properties of the specific model used. 
Which values are appropriate for the final-state flow coefficients is unknown. 
Therefore, our goal is not to correct for the effect of this correlation, but merely to assess its importance. 
For the sake of simplicity, we neglect the centrality dependence of $\alpha$ and $\alpha'_n$, and we choose constant values $\alpha=-2$, $\alpha'_2=\alpha'_3=-1$, indicated as dashed lines in Fig.~\ref{fig:correlations}. 
With these values, we repeat the fit to the minimum-bias simulations, shown as dash-dotted lines in Fig.~\ref{fig:cumulantsT}. 

The fit is slightly worse for the largest values of $N_{ch}$ (this is true in particular for $nc_2\{4\}$), which makes sense since they correspond to very small centralities, where values of $\alpha'_n$ closer to 0 would be preferred (Fig.~\ref{fig:correlations}). 
On the other hand, the values of $\bar\varepsilon_2(c)$ and $\sigma_{\varepsilon_2}(c)$ returned by the fit are in better agreement with the direct calculations, as shown by the dash-dotted lines in Fig.~\ref{fig:fixedbT}.  

We are now in a position to assess the robustness of the reconstruction carried out using ATLAS data in Sec.~\ref{s:data}. 
In Eqs.~(\ref{nongaussmoments}), we replace $\bar v$ and $\sigma_v^2$ with the expressions given in Eqs.~(\ref{alphaexp}) and (\ref{alphapexp}), and then repeat the fit of the data, using the aforementioned values $\alpha=-2$ and $\alpha'_2=\alpha'_3=-1$. 
In other words, we assume that the correlation between the final-state anisotropic flow, $V$, and the multiplicity, $N$, is the same as the correlation between initial-state anisotropy, $\varepsilon_n$ and the initial-state entropy, $S$, returned by the \trento{} model.
We redo the fit of $N_{ch}$-based ATLAS data (left panel of Fig.~\ref{fig:ATLAS}) on this basis. 
The resulting fit (not shown) underestimates $v_n\{2\}$ for the largest values of $N_{ch}$, which results in a larger chi square, as shown in Table~\ref{fitparameters}. 
The corrected reconstructed centrality dependence of $\bar v_2$ and $\sigma_{v_n}$ is displayed in Fig.~\ref{fig:fixedb} as dash-dotted lines. 
The effect of the correlation is qualitatively the same as in the \trento{} results of Fig.~\ref{fig:fixedbT}. 
It is a rather small effect. 
The conclusion is that, even though the correlation between anisotropic flow $V$ and the centrality estimator $N$ is unknown, the reconstruction of the centrality dependence from data is robust. 
It would be interesting to study this phenomenon in greater depth in future, by either performing state-of-the-art hydrodynamic calculations where the correlation between $V$ and $N$ can be explicitly evaluated, or by analyzing some appropriate final-state observable that may quantify this correlation.

\subsection{Non-Gaussian eccentricity fluctuations}
\label{s:nongaussianT}

\begin{figure}[t]
\begin{center}
\includegraphics[width=\linewidth]{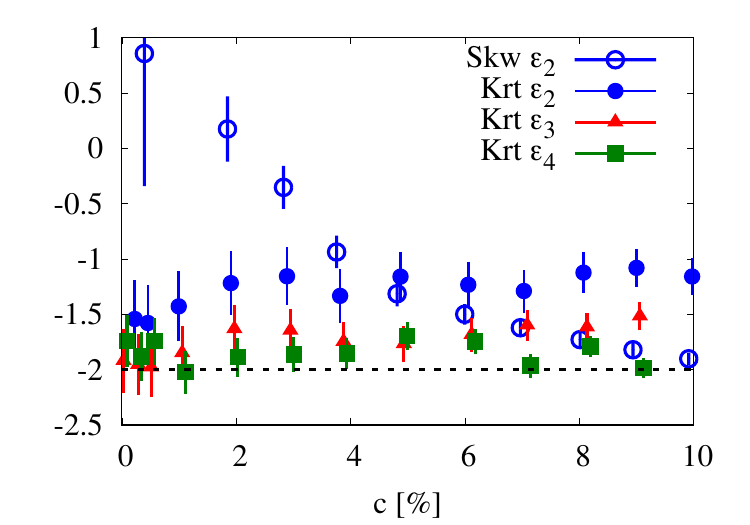} 
\end{center}
\caption{
Non-Gaussianity of initial anisotropy fluctuations at fixed centrality in the \trento{} model. 
We display the centrality dependence of the intensive skewness (for $\varepsilon_2$) and  kurtosis of $\varepsilon_{n}$ fluctuations, quantified by the coefficients ${\tt Skw}_\varepsilon$ and ${\tt Krt}_\varepsilon$ defined in Appendix~\ref{s:cumulants}. 
Vertical bars are estimates of statistical errors. 
Different symbol types have been slightly displaced horizontally to avoid overlap.  
The horizontal dotted line is the kurtosis value for the Power distribution~\cite{Yan:2014afa}. 
}
\label{fig:nongaussianT}
\end{figure}

We finally discuss the impact of the non-Gaussianity of the eccentricity fluctuations, characterized by their skewness and kurtosis at fixed centrality. 
Figure~\ref{fig:nongaussianT} displays the centrality dependence of the intensive skewness and kurtosis, defined by Eq.~(\ref{defintensive}) in Appendix~\ref{s:cumulants}. 
They are defined in such a way that they should depend little on centrality, and the values are indeed approximately constant. 
The kurtosis of $\varepsilon_3$ and $\varepsilon_4$ fluctuations is remarkably close to a value of $-2$, which corresponds to the value of the non-Gaussianities for the Power distribution~\cite{Yan:2013laa, Yan:2014afa}, in the limit of a large system ($\alpha\gg 1$). 
We have checked that this result is robust as one varies the fluctuation parameter, $k$, of the \trento{} calculations. 
We postulate that it is, to a large extent, a universal property. 
This should be checked in dedicated studies.

The next question is whether or not we are able to reconstruct this non-Gaussianity from the cumulants in Fig.~\ref{fig:cumulantsT}. 
${\tt Krt}_{\varepsilon_3}$ and ${\tt Krt}_{\varepsilon_4}$ are fairly well reconstructed.
If we assume that the distribution of $\varepsilon_n$ at fixed $c$ solely depends on centrality (dash-dotted lines in Fig.~\ref{fig:cumulantsT}), we obtain ${\tt Krt}_{\varepsilon_3}=-2.85\pm 0.02$ and ${\tt Krt}_{\varepsilon_4}=-2.69\pm 0.02$. 
If we introduce the correlation between $\varepsilon_n$ and multiplicity, i.e., initial-state entropy (solid lines in  Fig.~\ref{fig:cumulantsT}), we obtain ${\tt Krt}_{\varepsilon_3}=-1.94\pm 0.02$ and 
${\tt Krt}_{\varepsilon_4}=-1.91\pm 0.02$, in much better agreement with the direct calculation shown in Fig.~\ref{fig:nongaussianT}. 
For the second harmonic, the errors on ${\tt Skw}_{\varepsilon_2}$ and ${\tt Krt}_{\varepsilon_2}$ are an order of magnitude larger than for $\varepsilon_3$ and $\varepsilon_4$.   
These intrinsic non-Gaussianities are hidden by the mean eccentricity in the reaction plane.  

We have also studied how the skewness and kurtosis reconstructed from ATLAS data vary depending on whether or not one takes into account the correlation between $V$ and multiplicity. The results are given in Tab.~\ref{fitparameters} 
For $v_2$, they vary a lot and change sign, which further confirms that the intrinsic non-Gaussianity is poorly constrained. 
The kurtosis of $v_3$ also changes by more than a factor 2. 
That of $v_4$, on the other hand, is remarkably stable. 

\section{Conclusions}
\label{s:conclusions}
 
We have shown that the peculiar behavior exhibited by the cumulants of anisotropic flow fluctuations as they approach the limit of ultracentral collisions is a consequence of the finite centrality resolution, i.e., of impact parameter fluctuations at fixed values of the experimental centrality estimators. 
More specifically, we have shown that the fluctuations of $v_2$ at fixed impact parameter are essentially Gaussian, and that the positive value of $c_2\{4\}$ is a generic consequence of impact parameter fluctuations.

On the other hand, non-Gaussian fluctuations of $v_3$ and $v_4$ mostly reflect intrinsic non-Gaussanities. 
We have postulated that the intensive kurtosis of initial anisotropy fluctuations is universal to a good approximation. 
If this is correct, then the intensive kurtosis of $v_4$ fluctuations, which is accurately reconstructed, gives direct access to the hydrodynamic response coefficient, which is the ratio $v_4/\varepsilon_4$. 
This is the first data-driven estimate of this coefficient, which does not rely on a specific  model of initial conditions. 

We have also shown that the impact parameter dependence of anisotropic flow (specifically, of the mean elliptic flow in the reaction plane, and of $v_n$ fluctuations) can be reconstructed from data in a fairly robust way. 
This has the potential to greatly facilitate theory-to-data comparisons in future, as running hydrodynamic simulations at fixed impact parameter is straightforward, requiring limited statistics of events to reach a high accuracy.

One of the outputs of our reconstruction is that anisotropic flow fluctuations increase much faster as a function of impact parameter than one would expect from naive system-size scaling. 
This fast increase is, in particular, responsible for the large positive value of $nc_2\{4\}$ observed by ATLAS. 
Under the assumption that the hydrodynamic response is linear, this gives a new, direct data-driven constraint on models of initial conditions.

\begin{acknowledgments}
 M. Alqahtani acknowledges the support of the Research Mobility Program of the French Embassy in Riyadh. He also acknowledges that this work benefited from State aid under France 2030 (P2I -Graduate School Physique) bearing the reference ANR-11-IDEX-0003.
 G.G.~and A.K.~are funded by the Deutsche Forschungsgemeinschaft (DFG, German Research Foundation) – Project-ID 273811115 – SFB 1225 ISOQUANT, and under Germany's Excellence Strategy EXC2181/1-390900948 (the Heidelberg STRUCTURES Excellence Cluster).
\end{acknowledgments}

\appendix
\section{Cumulant expansion in the intrinsic frame}
\label{s:cumulants}

In this appendix, we define cumulants of fluctuations of initial anisotropy coefficients, $\varepsilon_n=(\varepsilon_x,\varepsilon_y)$. 
The same formalism can be applied to $V_n$ fluctuations. 
There are two differences between the cumulant expansion presented below and that done in experiments. 
The first difference is that we consider an ensemble of events with the same {\it centrality\/} $c$, as opposed to an ensemble of events with the same {\it centrality estimator\/}. This, however, does not modify the formalism. 
The second difference is that we work in the ``intrinsic frame'', where the impact parameter is along the $x$ axis. 
Except for $c=0$, where the ensemble of events possesses azimuthal symmetry, the cumulants defined below differ from the usual ones, and cannot be measured. 
But they can be evaluated in model calculations, where the impact parmeter is known. 

The reason why the cumulant expansion as carried out here is useful is that it can be organized systematically as a function of two key parameters: 
The first parameter is the system size, which we denote by $N$.\footnote{In the main text, $N$ denotes a slightly different quantity, namely, the centrality estimator.} 
The underlying picture is that initial fluctuations are in the form of $N$ independent sources~\cite{Bhalerao:2006tp} (whose order of magnitude is the number of participant nucleons~\cite{Miller:2007ri}), 
but the formalism is more general than this specific picture. 
The second parameter is the mean reaction plane eccentricity, whose definition will be recalled below, and which we denote by $\bar\varepsilon$. 
It vanishes for $c=0$. For $c>0$, it quantifies the magnitude of breaking of azimuthal symmetry. 

We derive the leading corrections to the Gaussian distribution assuming that the system is large ($N\gg 1$) and almost symmetric ($\bar\varepsilon\ll 1$), which is the appropriate limit for ultracentral nucleus-nucleus collisions. 

Let $(\varepsilon_x,\varepsilon_y)$ denote the initial anisotropy of an event in a given Fourier harmonic in the intrinsic frame. 
The corresponding cumulants $\kappa_{mp}$  were defined by Abbasi {\it et al.\/}~\cite{Abbasi:2017ajp} as a double series of connected moments (see also \cite{Bhalerao:2018anl,Giacalone:2024ixe}): 
\begin{equation}
  \label{kappamn}
\kappa_{mp}=\langle \varepsilon_x^m\varepsilon_y^p\rangle_c,
\end{equation}
where the subscript $c$ means that one takes the connected part of the product by subtracting out all the lower-order correlations, e.g.,
$\langle  \varepsilon_x^2\rangle_c\equiv\langle\varepsilon_x^2\rangle-\langle\varepsilon_x\rangle^2$.

When the system possesses azimuthal symmetry, either exact or approximate, simplifications occur, which appear more clearly if one introduces the complex eccentricity $\varepsilon=\varepsilon_x+i\varepsilon_y$~\cite{Qiu:2011iv}. 
Instead of the cumulants of $\varepsilon_x$ and $\varepsilon_y$, we introduce the cumulants of $\varepsilon$ and the complex conjugate $\varepsilon^*$, which are defined through the following generating function of a complex parameter $\lambda$: 
\begin{eqnarray}
\label{gencumulants}
\ln\left\langle \exp\left(\lambda\varepsilon^*+\lambda^*\varepsilon\right)\right\rangle
=\sum_{m,p\ge 0}\frac{\lambda^m\lambda^{*p}}{m!p!}c_{mp},
\end{eqnarray}
where angular brackets denote an ensemble average over events at fixed centrality $c$. 
Note that $c_{00}=0$. 

For $m+p\ge 1$, the cumulants $c_{mp}$ defined by Eq.~(\ref{gencumulants}) are connected moments of the complex eccentricity and its conjugate: 
\begin{equation}
  \label{defcmn}
c_{mp}=\langle \varepsilon^m(\varepsilon^{*})^p\rangle_c.
\end{equation}
Symmetry with respect to the $x$ axis implies that $c_{mp}$ is real and symmetric, $c_{mp}=c_{pm}$. 
If the distribution is azimuthally symmetric, then only the diagonal terms $c_{mm}$ are non-vanishing. 

We will assume that the distribution of $\varepsilon$ is azimuthally symmetric for the third and fourth Fourier harmonics. 
For the second Fourier harmonic, azimuthal symmetry is broken by the reaction plane eccentricity $\bar\varepsilon\equiv c_{10}$, which is treated as a small parameter. 
The order of magnitude of the cumulants is generically:
\begin{equation}
\label{orderofmagnitude}
c_{mp}\sim \bar\varepsilon^{|m-p|} N^{1-m-p}. 
\end{equation}
For a fixed order $m+p$, they are strongly ordered as a function of the difference $|m-p|$, that is, $|c_{20}|\ll |c_{11}|$, $|c_{30}|\ll |c_{21}|$,
$|c_{40}|\ll |c_{31}| \ll |c_{22}|$.

The cumulants we shall need are the following: 
\begin{eqnarray}
\label{cumulantmatrix}
  \begin{pmatrix}
   c_{00}&c_{10}&c_{20}\cr
      &c_{11}&c_{21}\cr
      &&c_{22}
      \end{pmatrix}
  &=&  \begin{pmatrix}
    0&{\rm mean\ ecc.}&{\rm fluct.\ asymmetry}\cr
      &{\rm variance}&{\rm skewness}\cr
      &&{\rm kurtosis}
  \end{pmatrix}\cr
  &\sim&
  \begin{pmatrix}
    0&\bar\varepsilon&\bar\varepsilon^2/N\cr
    &1/N&\bar\varepsilon/N^2\cr
    &&1/N^3
  \end{pmatrix}    
\end{eqnarray}
The leading terms are $c_{10}=\bar\varepsilon$ and $c_{11}=\sigma_\varepsilon^2$ (mean eccentricity and variance of fluctuations) which define the usual Gaussian distribution of eccentricity~\cite{Voloshin:2007pc}.
We will include the skewness and kurtosis, which are smaller than $\bar\varepsilon$ and $\sigma_\varepsilon^2$ by $1/N^2$, and are the leading non-Gaussian correction, but we shall neglect the fluctuation asymmetry $c_{20}$ because it is of higher order in $\bar\varepsilon$, and negligible for small impact parameters. This justifies the ansatz in Eq.~(\ref{gaussianv}), where the standard deviations of $v_x$ and $v_y$ are assumed to be equal.

In order to suppress the dependence of the skewness and kurtosis on centrality, we introduce {\it intensive\/} measures~\cite{Giacalone:2020lbm}, which we denote by ${\tt Skw}_\varepsilon$ and ${\tt Krt}_\varepsilon$. 
They are related to the cumulants through the relations: 
\begin{eqnarray}
\label{defintensive}
c_{21}&=&\bar\varepsilon(\sigma_\varepsilon^2)^2 {\tt Skw}_\varepsilon\cr
c_{22}&=&(\sigma_\varepsilon^2)^3 {\tt Krt}_\varepsilon.
\end{eqnarray}
Eq.~(\ref{orderofmagnitude}) shows that ${\tt Skw}$ and ${\tt Krt}$ are of order unity. 
With these notations and approximations, the matrix of cumulants (\ref{cumulantmatrix}) becomes: 
\begin{equation}
\label{cumulantmatrix2}
  \begin{pmatrix}
   c_{00}&c_{10}&c_{20}\cr
      &c_{11}&c_{21}\cr
      &&c_{22}
      \end{pmatrix}
  =  \begin{pmatrix}
    0&\bar\varepsilon
   &0\cr
      &\sigma_\varepsilon^2&
      \bar\varepsilon(\sigma_\varepsilon^2)^2 {\tt Skw}_\varepsilon
      \cr
      &&(\sigma_\varepsilon^2)^3 {\tt Krt}_\varepsilon
  \end{pmatrix}.
\end{equation}
Since the cumulants $\kappa_{mp}$ in Eq.~(\ref{kappamn}) have been used previously in the literature, we list their relations with the cumulants appearing in Eq.~(\ref{cumulantmatrix2}): 
\begin{eqnarray}
c_{10}&=&\kappa_{10}\cr
c_{11}&=&\kappa_{20}+\kappa_{02}\simeq 2\kappa_{20} \cr
c_{20}&=&\kappa_{20}-\kappa_{02}\cr
c_{21}&=&\kappa_{30}+\kappa_{12}\simeq \frac{4}{3}\kappa_{30}\cr
c_{22}&=&\kappa_{40}+\kappa_{04}+2\kappa_{22}\simeq  \frac{8}{3}\kappa_{40},
\end{eqnarray}
where the approximate equalities have been obtained by neglecting terms of order $\bar\varepsilon^2$ and higher, which amounts, using Eq.~(\ref{orderofmagnitude}), to neglecting $c_{nm}\approx 0$ for $|m-p|> 1$, specifically $c_{20}=c_{30}=c_{31}=c_{40}=0$. 
  
The moment of order $2k$,   $\langle |\varepsilon_n|^{2k}\rangle$, which is needed in Sec.~\ref{s:nongaussian}, is obtained by exponentiating Eq.~(\ref{gencumulants}), expanding in a double series of $\lambda$ and $\lambda^*$, and isolating the coefficient of $(\lambda\lambda^*)^k$ in the power-series expansion, which is straightforward using formal calculus. 
This is how the explicit expressions (\ref{nongaussmoments}) are obtained.  

Finally, the cumulants of the correlations between the initial anisotropy and the total entropy (which we denote here by $N$ since the final multiplicity is  proportional to the entropy, to a good approximation), which are used in Sec.~\ref{s:correlations},  are given by a straightforward generalization of Eq.~(\ref{gencumulants}): 
\begin{equation}
\label{gencumulantsN}
\ln\left\langle \exp\left(\lambda\varepsilon^*+\lambda^*\varepsilon+\mu N\right)\right\rangle
=\sum_{m,p\ge 0}\frac{\lambda^m\lambda^{*p}\mu^q}{m!p!q!}c_{mpq}. 
\end{equation}
With this notation, the linear correlation between $\varepsilon_2$ and $N$ (left-hand side of Eq.~(\ref{lincorr})) is $c_{101}$. 
We also provide, for the sake of completeness, the explicit expression of $c_{111}$, which is necessary in order to evaluate the exponent $\alpha'_n$ in Eq.~(\ref{alphapexp}): 
\begin{align}
\label{c111}
\nonumber c_{111}&= \langle \varepsilon\varepsilon^*N\rangle_c \\
&= \langle \varepsilon\varepsilon^*N\rangle-\langle \varepsilon\varepsilon^*\rangle\langle N\rangle- 2 \langle \varepsilon N\rangle\langle\varepsilon\rangle+2\langle \varepsilon \rangle^2 \langle N\rangle.
\end{align}
In the same way, one could evaluate the correlations of the skewness and kurtosis with the multiplicity, which correspond respectively to $c_{211}$ and $c_{221}$. 

\bibliography{ultracentral_vn}

\end{document}